\documentclass[11pt,a4paper]{article}                           

\usepackage{amsmath}
\usepackage{amssymb}
\usepackage[usenames]{color} 
\usepackage{multirow}
\usepackage{graphicx}
\usepackage{epstopdf}
\usepackage{array}
\usepackage{hhline}
\usepackage{cite}
\usepackage{microtype,colortbl}
\usepackage{color}
\usepackage[usenames,dvipsnames]{xcolor}
\usepackage{pstool}
\usepackage{tikz}
\usepackage{ytableau}
\usepackage{latexsym,stmaryrd}
\usepackage[bookmarks=false]{hyperref}
 \usepackage{lscape}
\usepackage{caption}
 \usepackage{subcaption}
 \usepackage{tensor}
 \usepackage{dsfont}
\usepackage{ifpdf}
\def\hybrid{\topmargin -20pt    \oddsidemargin 0pt
        \headheight 0pt \headsep 0pt
        \textwidth 6.25in       % A4 paper
        \textheight 9 in       % A4 paper
        \marginparwidth .875in
        \parskip 5pt plus 1pt 
          \jot = 1.5ex
   }
\hybrid
\numberwithin{equation}{section}
\numberwithin{table}{section}

\newcommand{\beq}{\begin{equation}}  \newcommand{\eeq}{\end{equation}}
\newcommand{\bal}{\begin{aligned}}   \newcommand{\eal}{\end{aligned}}
\newcommand{\bea}{\begin{eqnarray}}  \newcommand{\eea}{\end{eqnarray}}

\newcommand{\bmat}{\left(\begin{array}}
\newcommand{\emat}{\end{array}\right)}

\newcommand{\bbR}{\mathbb{R}}

\usepackage{cancel}

\newcommand{\be}{\begin{equation}}
\newcommand{\ee}{\end{equation}}

\DeclareMathOperator\arctanh{arctanh}
\DeclareMathOperator{\arccot}{arccot}

\definecolor{Gray}{gray}{0.95}

\begin{document}
\baselineskip=14pt
\parskip 5pt plus 1pt 

\vspace*{3cm}
\begin{center}
	{\LARGE\bfseries Bi-Yang--Baxter Models and Sl(2)-orbits}\\[.3cm]
	
	\vspace{1.5cm}
	{\bf Thomas W.~Grimm}\footnote{t.w.grimm@uu.nl},
	{\bf Jeroen Monnee}\footnote{j.monnee@uu.nl}
	
	{\small
		\vspace*{.5cm}
		Institute for Theoretical Physics, Utrecht University\\ Princetonplein 5, 3584 CC Utrecht, The Netherlands\\[3mm]
	}
\end{center}
\vspace{3cm}
\begin{abstract}\noindent
We study integrable deformations of two-dimensional non-linear $\sigma$-models and present a new class of classical solutions to  critical bi-Yang--Baxter models for general groups. For the simplest example, namely the $\mathrm{SL}(2,\mathbb{R})$ bi-Yang--Baxter model, we show that our solutions can be mapped to the known complex uniton solutions of the $\mathrm{SU}(2)$ bi-Yang--Baxter model.
In general, our solutions are constructed from so-called $\mathrm{Sl}(2)$-orbits that play a central role in the study of asymptotic Hodge theory. This provides further evidence for a close relation between integrable non-linear $\sigma$-models and the mathematical principles underlying Hodge theory. We have also included a basic introduction to the relevant aspects of asymptotic Hodge theory and have provided some simple examples. 
\end{abstract}

\newpage

\tableofcontents
\setcounter{footnote}{0}

\newpage
%%%%%%%%%%%%%%%%%%%%%%%%%%%%%%%%%%%%%%%%%%%%%%%%%%%%%%

\section{Introduction}

In general quantum field theories the computation of correlation functions is notoriously difficult and can often only be performed in perturbation theory. The situation improves significantly if  symmetries constrain the theory. A particularly interesting class of models, 
known as integrable models, are constrained by having infinitely many conserved currents. Despite the fact that integrability poses a strong condition on the theory, a large set of integrable models have been constructed in the past decades \cite{Hoare:2021dix,Driezen:2021cpd}. In this work we will focus on a particular 
class of two-dimensional integrable non-linear $\sigma$-models and use some of the powerful tools of Hodge theory
in determining their classical solutions. Our finding will further strengthen the connection between integrable models and this vast field of mathematics first observed in \cite{Grimm:2021idu}.

The class of integrable models considered here are known as bi-Yang--Baxter models, which were introduced in \cite{Klimcik:2008eq}. To construct these models one starts with the principal chiral model, which is a
non-linear $\sigma$-model encoding the dynamics of a field valued in a group $G$. The bi-Yang--Baxter model is then defined as a two-parameter deformation that depends on a Yang--Baxter operator $R$ satisfying a modified classical Yang--Baxter equation. These models can be abstractly defined for any group $G$ when making sure that an appropriate $R$-matrix is constructed, e.g.~following the classical work of Drinfel'd--Jimbo \cite{Drinfeld:466366,Jimbo:466362}, see also \cite{Belavin:1982}. Despite its general definition, the study of solutions of such models has so far been restricted to only the simplest choices of $G$. In particular, the $G=\mathrm{SU}(2)$ bi-Yang--Baxter model has been investigated in \cite{Schepers:2020ehn}. Our first aim is to provide a new perspective on the classical 
solutions found in \cite{Schepers:2020ehn} that allows for a natural generalization to higher-rank groups $G$. 

Our study of solutions to the bi-Yang--Baxter model will be restricted to a special one-parameter subspace of the two-parameter moduli space, where the symmetries of the model enhance and additional dualities to other theories emerge \cite{Klimcik:2008eq,Klimcik:2014bta,Delduc:2013qra,Hoare:2014oua}. 
These models will be referred to as critical bi-Yang--Baxter models following \cite{Schepers:2020ehn}. In the simplest 
situation, namely the critical SL$(2,\bbR)$ bi-Yang--Baxter model, we will show that the equations of motion are solved by the so-called Weil operator $C$, which is sometimes also referred to as the Hodge star, 
associated to a two-torus. This operator distinguishes complex $(1,0)$-forms and $(0,1)$-forms and changes 
upon varying the complex structure of the two-torus. 
Hence, to view $C$ as a solution to the $\sigma$-model requires us to identify the complex structure deformation space of the two-torus, namely the upper half-plane, with part of the two-dimensional space-time.\footnote{This shift in perspective is key to all approaches \cite{Cecotti:2020uek,Grimm:2020cda,Grimm:2021ikg,Grimm:2021idu} that use Hodge theory to construct solutions to two-dimensional $\sigma$-models.} Furthermore, it is a general fact that the Weil operator of a two-torus obeys $C^2=-1$, and hence that the solutions must be a special class of all possible solutions. Exactly these types of solutions were called uniton solutions in the literature on integrable models and were first considered for the SU$(N)$ principal chiral model in \cite{Uhlenbeck:1989}. We explicitly show 
that the Weil operator is equivalent to a certain complex uniton solution of the $\mathrm{SU}(2)$ bi-Yang--Baxter
model that has been constructed in \cite{Schepers:2020ehn}, see also \cite{Demulder:2016mja}. 

A careful inspection of the Weil operator of a two-torus reveals an intriguing underlying structure, namely that it can be written as an $\mathrm{Sl}(2)$-orbit. Roughly speaking, this means that the Weil operator changing over space-time can be written as an orbit of some special fixed Weil operator. The orbit is derived by picking 
a distinguished element of $\mathrm{SL}(2,\mathbb{R})$ and parameterizing the transformations by this element with a complex parameter labelling the space-time position. In the case of the two-torus the appearance of such orbits might not be surprising, but remarkably it turns out that this perspective can be generalized to more general groups and that $\mathrm{Sl}(2)$-orbits are in fact ubiquitous in the study of asymptotic Hodge theory. In practice, it is relatively straightforward to construct such $\mathrm{Sl}(2)$-orbits and the relevant ingredients are captured by a so-called horizontal $\mathfrak{sl}(2)$-triple. This is simply a standard $\mathfrak{sl}(2)$-triple, i.e.~three real elements $N^+$, $N^-$, and $N^0$ in the algebra of $G$ obeying $\mathfrak{sl}(2)$ commutation relations, that satisfies some additional commutation relations with respect to another operator $Q_\infty$. The latter introduces an additional grading on the algebra of $G$ and has the property $\bar Q_\infty = - Q_\infty$. Given such a horizontal $\mathfrak{sl}(2)$-triple, we argue that it selects a particular class of $R$-matrices and we explicitly show that the associated $\mathrm{Sl}(2)$-orbit solves the equations of motion of the corresponding critical bi-Yang--Baxter model. In general, the resulting solution satisfies $C^2=(-1)^D$, for some integer $D$, and furthermore has finite action. It can therefore be thought of as a generalization of the complex uniton solution of the $\mathrm{SL}(2,\mathbb{R})$ model to groups of higher rank. For illustrative purposes, we have included an explicit example of a solution to the critical  $\mathrm{Sp}(4,\mathbb{R})$ bi-Yang--Baxter model obtained in this manner.

The mathematical significance of $\mathrm{Sl}(2)$-orbits is rooted in one of the deep theorems of Hodge theory, the $\mathrm{Sl}(2)$-orbit theorem \cite{schmid,CKS}. Very roughly, it states that whenever one has a Hodge structure varying over some parameter space, the behaviour of this variation of Hodge structure near the boundary of the parameter space is described by an $\mathrm{Sl}(2)$-orbit. As mentioned earlier, such 
$\mathrm{Sl}(2)$-orbits can be structured in terms of horizontal $\mathfrak{sl}(2)$-triples, and the latter have been classified in the mathematics literature \cite{Robles:2015,Kerr2017}. Therefore, these general mathematical results provide a concrete classification of uniton solutions of the bi-Yang--Baxter model and further indicates the significance of Hodge theory in the study of integrable models. In this work, we do not aim to develop this latter point of view in full, but rather aim to lay the foundations for further studying this connection. To this end, we provide a comprehensive introduction to the relevant aspects of asymptotic Hodge theory and illustrate the central ideas with a simple example. 

The paper is organized as follows. In section \ref{sec:bi_YB_unitons} we introduce the bi-Yang--Baxter model. The action of this theory depends on the choice of an $R$-matrix and we recall how the Drinfel'd--Jimbo solution for $R$ indeed satisfies the classical modified Yang-Baxter equation. We then discuss the SU$(2)$ example in detail and introduce its uniton solutions. We give a detailed account on how this solution can be related to the Weil operator of a two-torus and point out that the solution can be described purely in terms of a horizontal $\mathfrak{sl}(2)$-triple. In section \ref{sec:Hodge_Theory} the solution is then extended to higher-rank groups by identifying the group-valued field with the general $\mathrm{Sl}(2)$-orbit of special fixed Weil operators. We argue that the horizontal $\mathfrak{sl}(2)$-triple selects a particular class of $R$-matrices and explicitly show that the equations of motion of the corresponding bi-Yang--Baxter model are solved at the critical point. We also provide an introduction to the relevant aspects of asymptotic Hodge theory to explain the importance of $\mathrm{Sl}(2)$-orbits and horizontal $\mathfrak{sl}(2)$-triples. We end with an illustrative example and comment on possible generalizations of the proposed solutions. In the appendix we have included some computational details and elaborate on some of the expressions used in the example of section~\ref{subsec:example}.
\pagebreak

\section{The Bi-Yang--Baxter Model and Unitons}
\label{sec:bi_YB_unitons}

In this section we analyze classical aspects of the bi-Yang--Baxter model. In section \ref{subsec:bi_YB} we introduce the model and establish our notation and conventions. Then, in section \ref{subsec:unitons}, we study the $\mathrm{SU}(2)$ model in more detail and consider a class of finite action solutions known as \textit{unitons}. Finally, in section \ref{subsec:Weil_SL2} we observe a relation between the complex uniton solution and the Weil operator of a two-torus. This observation will then lead us to consider more general solutions for Weil operators of arbitrary variations of Hodge structure in section \ref{sec:Hodge_Theory}. 

\subsection{Bi-Yang--Baxter Model}
\label{subsec:bi_YB}

Let us start by introducing the basics of the bi-Yang--Baxter model. The reader who is already familiar with the topic can safely skip this section. The model was originally introduced by C.~Klim\v{c}\'ik in \cite{Klimcik:2008eq} as a two-parameter integrable deformation of the principal chiral model. In particular, it is a non-linear $\sigma$-model for a group-valued field
\begin{equation}
    g:\Sigma\rightarrow G\,,
\end{equation}
where $\Sigma$ is the two-dimensional worldsheet and $G$ is a real Lie group, whose Lie algebra will be denoted by $\mathfrak{g}$. We will take the worldsheet to have Euclidean signature, and introduce complex coordinates $z,\bar{z}$ with $z=x+iy$. Additionally, we will assume $\mathfrak{g}$ to be simple and denote by 
\begin{equation}
    (\cdot,\cdot):\mathfrak{g}\times\mathfrak{g}\rightarrow\mathbb{R}
\end{equation}
the (up to an overall scaling) unique invariant symmetric bilinear form on $\mathfrak{g}$.

\subsubsection*{$R$-matrix}
The bi-Yang--Baxter model lies in the class of Yang--Baxter deformations of non-linear $\sigma$-models, all of which involve an object called the (classical) $R$-matrix. It is an endomorphism of the Lie algebra $\mathfrak{g}$, i.e.~a linear map
\begin{equation}
    R:\mathfrak{g}\rightarrow\mathfrak{g}\,,
\end{equation}
satisfying the modified classical Yang--Baxter equation
\begin{equation}
\label{eq:YBE}
    [RX, RY]-R\left([RX, Y]+[X,RY]\right)=-c^2[X,Y]\,,\qquad \forall X,Y\in\mathfrak{g}\,,
\end{equation}
where $c^2$ is a real constant. The word `modified' refers to the fact that $c$ is allowed to be non-zero. Note that by a real rescaling of $R$ we may restrict to the cases $c\in\{0,1,i\}$. In the following, we will additionally impose the condition that $R$ is skew-symmetric with respect to the chosen bilinear form on $\mathfrak{g}$. In other words
\begin{equation}
\label{eq:skew}
    (RX,Y)+(X,RY)=0\,,\qquad \forall X,Y\in\mathfrak{g}\,.
\end{equation}
Such $R$-matrices are also referred to as Yang--Baxter operators and have been classified in the mathematics literature \cite{Belavin:1982}, see also \cite{Chari:1995}. 

\subsubsection*{Drinfel'd--Jimbo $R$-matrix}

There is a standard solution to the modified classical Yang--Baxter equation, given by the Drinfel'd--Jimbo solution \cite{Drinfeld:466366,Jimbo:466362}. In order to write it down, let us first consider the complexification $\mathfrak{g}_\mathbb{C}$ of $\mathfrak{g}$ and let $\mathfrak{h}$ be a Cartan subalgebra of $\mathfrak{g}_\mathbb{C}$. Then $\mathfrak{g}_\mathbb{C}$ enjoys a root space decomposition
\begin{equation}
    \mathfrak{g}_\mathbb{C}=\mathfrak{h}\oplus\bigoplus_\alpha\mathfrak{g}_\alpha\,,
\end{equation}
where $\alpha\in\mathfrak{h}^*$ runs over all the roots and each $\mathfrak{g}_\alpha$ denotes the root space associated to a root $\alpha$. A choice of simple roots fixes a notion of positive roots, denoted by $\alpha>0$. Let $\{H^\mu, E^{\pm\alpha}\}$ denote a Cartan--Weyl basis of $\mathfrak{g}_\mathbb{C}$, where $\alpha$ runs over the positive roots. These generators satisfy the usual commutation relations
\begin{equation}
    [H^\mu, H^\nu]=0\,,\qquad [H^\mu, E^{\pm\alpha}]=\pm\alpha(H^\mu)E^{\pm\alpha}\,.
\end{equation}
In terms of this basis, the Drinfel'd--Jimbo $R$-matrix is defined as\footnote{The overall sign of $R$ is, of course, a matter of convention. Here we have chosen the sign to match with \cite{Hoare:2021dix}.}
\begin{equation}
\label{eq:def_R_DJ}
    RH^\mu=0\,,\qquad RE^{\pm\alpha}=\mp c\,E^{\pm\alpha}\,.
\end{equation}
One can verify by direct computation that \eqref{eq:def_R_DJ} solves the modified classical Yang--Baxter equation \eqref{eq:YBE} and that it satisfies the skew-symmetry condition \eqref{eq:skew}. However, it is important to keep in mind that the above $R$-matrix is defined on the complexified algebra $\mathfrak{g}_\mathbb{C}$. Depending on the choice of real form $\mathfrak{g}$ and the constant $c$, it may happen that $R$ is not a real endomorphism of $\mathfrak{g}$. For further details we refer the reader to the lecture notes of B.~Hoare \cite{Hoare:2021dix}.

\subsubsection*{Action}
For a given choice of Yang--Baxter operator $R$, the action of the associated bi-Yang--Baxter model is given by
\begin{equation}
\label{eq:biYB}
    S = \int_\Sigma d^2\sigma\,\left(g^{-1}\partial_+g\,,\frac{1}{1-\eta R-\zeta R^g}g^{-1}\partial_- g\right)\,,
\end{equation}
where $\eta$ and $\zeta$ are two constants parametrizing the deformation, and we have introduced the notation
\begin{equation}
    R^g:= \mathrm{Ad}_{g^{-1}}\circ R\circ\mathrm{Ad}_g\,.
\end{equation}
Clearly, for $\eta=\zeta=0$ one recovers the action of the principal chiral model, which enjoys a global $G_L\times G_R$ symmetry. Keeping $\zeta=0$ but letting $\eta$ be non-zero, the introduction of the operator $R$ breaks the global $G_R$ symmetry down to the $\mathrm{U}(1)_R^{\mathrm{rk}\,G}$ subgroup. In this case, one retrieves the action of the (single-parameter) Yang--Baxter model, which is also referred to as the $\eta$-model \cite{Klimcik:2002zj}. Upon letting $\zeta$ be non-zero also the global $G_L$ symmetry is broken down to the $\mathrm{U}(1)_L^{\mathrm{rk}\,G}$ subgroup. 

There is a special point in the parameter space where the symmetry of the bi-Yang--Baxter model is enhanced \cite{Klimcik:2014bta}. Indeed, whenever $\zeta=\eta$, which we will refer to as the \textit{critical line} following \cite{Schepers:2020ehn}, one has an additional symmetry given by $g\mapsto g^{-1}$. The critical bi-Yang--Baxter model will play a crucial role in this work, as we will find a set of solutions to the model which solve the model precisely when $\zeta=\eta$. There have been a number of observations that indicate that the critical bi-Yang--Baxter model can be related to other integrable models. For example, the critical $\mathrm{SU}(2)$ bi-Yang--Baxter model is equivalent to the coset $\mathrm{SO}(4)/\mathrm{SO}(3)$ $\eta$-model \cite{Delduc:2013qra,Hoare:2014oua}. Furthermore, at the conformal point $\zeta=\eta=\frac{i}{2}$ the target space geometry coincides with that of the $\mathrm{SU}(1,1)/\mathrm{U(1)}$ gauged WZW model with an additional $\mathrm{U}(1)$ boson \cite{Hoare:2014oua}.

\subsubsection*{Equations of Motion}
In order to write down the equations of motion of the bi-Yang--Baxter model, it is convenient to introduce the currents (we follow the conventions of \cite{Klimcik:2014bta})
\begin{equation}
\label{eq:def_Jpm}
    J_\pm= \mp\frac{1}{1\pm \eta R\pm \zeta R^g}j_\pm\,,\qquad j_\mu = g^{-1}\partial_\mu g\,.
\end{equation}
In terms of $J_\pm$, the equations of motion read
\begin{equation}
\label{eq:eom_biYB}
    \partial_+ J_--\partial_-J_+ - \eta[J_+,J_-]_R=0\,,
\end{equation}
where
\begin{equation}
\label{eq:R_bracket}
    [X,Y]_R:=[RX,Y]+[X,RY]\,,\qquad \forall X,Y\in\mathfrak{g}\,,
\end{equation}
defines a second Lie-bracket on $\mathfrak{g}$, by virtue of the classical Yang--Baxter equation \eqref{eq:YBE}. It is referred to as the $R$-bracket and is central for the underlying Poisson-Lie symmetry of the bi-Yang--Baxter model, see e.g.~\cite{Sfetsos:2015nya}.  

\subsubsection*{Integrability and Lax Connection}
It was shown by C.~Klim\v{c}\'ik in \cite{Klimcik:2014bta} that the bi-Yang--Baxter model is classically integrable. Vital in this regard is the condition that $R$ satisfies the modified classical Yang--Baxter \eqref{eq:YBE} equation and is anti-symmetric. The integrability condition means that the equations of motion of the bi-Yang--Baxter model can be reformulated as the zero-curvature condition of a Lax connection
\begin{equation}
    \mathcal{L}_\pm(\lambda) = \left(\eta(R-i)+\frac{2i\eta\pm(1-\eta^2+\zeta^2)}{1\pm \lambda}\right)J_\pm\,,
\end{equation}
where $\lambda\in\mathbb{C}$ is the spectral parameter. More precisely, introducing the connection
\begin{equation}
    \nabla_\pm = \partial_\pm+\mathcal{L}_\pm(\lambda)\,,
\end{equation}
the zero-curvature condition simply states that
\begin{equation}
    [\nabla_+,\nabla_-]=0\,,
\end{equation}
for all $\lambda$. This is equivalent to the equations of motion \eqref{eq:eom_biYB} together with the Bianchi identities for $J_\pm$. The flatness of the Lax connection ensures the existence of an infinite tower of conserved charges. The Hamiltonian integrability of the model, i.e. the condition that all these charges in fact Poisson-commute with each other, was established in \cite{Delduc:2015xdm}.

\subsection{$\mathrm{SU}(2)$ unitons}
\label{subsec:unitons}

In this section, we restrict to the case where $G=\mathrm{SU}(2)$ and study a class of finite action solutions to the classical theory. These solutions are referred to as \textit{unitons}, owing to the fact that they are analogous to instantons and additionally satisfy the requirement that $g^2=-1$. They were originally constructed by K.~Uhlenbeck as solutions to the $\mathrm{SU}(N)$ principal chiral model \cite{Uhlenbeck:1989}, and were later extended to the Yang--Baxter and bi-Yang--Baxter models in \cite{Demulder:2016mja,Schepers:2020ehn} for $N=2$. Our discussion closely follows the works \cite{Demulder:2016mja,Schepers:2020ehn}.

\subsubsection*{$\mathfrak{su}(2)$ $R$-matrix}

For $\mathfrak{su}(2)$, the solution to the modified Yang--Baxter equation is essentially unique and is given by the Drinfel'd--Jimbo solution discussed in the previous section. Let us go through the construction of the $R$-matrix in some detail. The complexification of $\mathfrak{su}(2)$ is $\mathfrak{sl}(2,\mathbb{C})$, which has a Cartan--Weyl basis given by
\begin{equation}
    H = \begin{pmatrix}
    1&0\\
    0&-1
    \end{pmatrix}\,,\qquad E_+ = \begin{pmatrix}
    0&1\\
    0&0
    \end{pmatrix}\,,\qquad E_- = \begin{pmatrix}
    0&0\\
    1&0
    \end{pmatrix}\,,
\end{equation}
satisfying
\begin{equation}
    [H,E_\pm]=\pm 2 E_\pm\,.
\end{equation}
Using \eqref{eq:def_R_DJ} we obtain a solution to the modified classical Yang--Baxter equation, at the level of $\mathfrak{sl}(2,\mathbb{C})$. In order to see if this descends to a real solution when restricting to the real form $\mathfrak{su}(2)$, let us fix a basis of $\mathfrak{su}(2)$ as $T_j=i\sigma_j$, where $\sigma_i$ denote the Pauli matrices. Then one readily finds that $R$ acts on this basis as
\begin{equation}
    R T_1=-ic\,T_2\,,\qquad R T_2 = ic\, T_1\,,\qquad RT_3=0\,. 
\end{equation}
In particular, for $R$ to be a real endomorphism we require $c=i$, in which case $R$ can be represented as a matrix in the $T_i$ basis as
\begin{equation}
\label{eq:R_su2}
    R = \begin{pmatrix}
    0&-1&0\\
    1&0&0\\
    0&0&0
    \end{pmatrix}\,.
\end{equation}
In the remainder of this section, we will implicitly use this $R$-matrix. 

\subsubsection*{$\mathrm{SU}(2)$ bi-Yang--Baxter Model}
We will adopt the following parametrization of the $\mathrm{SU}(2)$ group element
\begin{equation}
\label{eq:g_SU2}
    g = \begin{pmatrix}
    \cos\theta\,e^{i\phi_1} & i\sin\theta\,e^{i\phi_2}\\
    i\sin\theta\,e^{-i\phi_2} & \cos\theta\,e^{-i\phi_1}
    \end{pmatrix}\,,\qquad \theta,\phi_1\in[0,\pi)\,,\quad \phi_2\in[0,2\pi)\,.
\end{equation}
As a non-linear $\sigma$-model, the bi-Yang--Baxter model can be characterized by the metric and $B$-field it induces on the target space. These follow from inserting the ansatz \eqref{eq:g_SU2} for $g$, together with the $R$-matrix \eqref{eq:R_su2}, into the action \eqref{eq:biYB}. The resulting metric reads
\begin{align*}
    ds^2&=\frac{1}{\Delta}\left[\mathrm{d}\theta^2+\cos^2\theta\left(1+(\eta+\zeta)^2\cos^2\theta\right)\mathrm{d}\phi_1^2+\sin^2\theta\left(1+(\eta-\zeta)^2\sin^2\theta\right)\mathrm{d}\phi_2^2\right]\\
    &\qquad +\frac{\sin^2(2\theta)}{2\Delta}(\eta-\zeta)(\eta+\zeta)\mathrm{d}\phi_1\mathrm{d}\phi_2\,,
\end{align*}
where we have defined
\begin{equation}
    \Delta = 1+\eta^2+\zeta^2+2\eta\zeta\cos(2\theta)\,.
\end{equation}
Moving on, the $B$-field is found to be pure-gauge, and given is by
\begin{equation}
    B=\mathrm{d}A\,,\qquad A=\frac{\log\Delta}{4\eta\zeta}\left[(\eta-\zeta)\mathrm{d}\phi_1-(\eta+\zeta)\mathrm{d}\phi_2\right]\,.
\end{equation}
The form of the metric and $B$-field indicate that at the points $\zeta=\eta$ and $\zeta=-\eta$ the model simplifies significantly. We recall that the point $\zeta=\eta$ is referred to as the critical line. In contrast, the point $\zeta=-\eta$ is referred to as the co-critical line. As alluded to before, on the critical line this simplification is due to the emergence of the $\mathbb{Z}_2$ symmetry $g\mapsto g^{-1}$. In the parametrization \eqref{eq:g_SU2} this corresponds to
\begin{equation}
    \phi_1\mapsto -\phi_1\,,\qquad \phi_2\mapsto \phi_2+\pi\,.
\end{equation}
However, note that for the $\mathrm{SU}(2)$ model in particular there is an additional $\mathbb{Z}_2$-symmetry given by the transformations
\begin{equation}
\label{eq_Z2_cocritical}
    \theta\mapsto \theta+\frac{\pi}{2}\,,\qquad \phi_1\leftrightarrow\phi_2\,,\qquad \zeta\mapsto -\zeta\,.
\end{equation}
This exactly maps the critical line $\zeta=\eta$ to the co-critical line $\zeta=-\eta$.

\subsubsection*{Real and Complex Unitons}
The unitons are finite action solutions to the classical equations of motion of the $\mathrm{SU}(2)$ bi-Yang--Baxter model, which additionally satisfy $g^2=-1$. In the parametrization \eqref{eq:g_SU2}, this condition imposes that either $\theta=\frac{\pi}{2}$ or $\phi_1=\frac{\pi}{2}$. We will consider the latter case. There are two types of unitons, dubbed the real and complex unitons. Both are determined by the choice of a holomorphic function $f(z)$ of the worldsheet coordinate. The expressions for $\phi_1$ and $\phi_2$ are the same for both unitons, and are given by
\begin{equation}
    \phi_1=\frac{\pi}{2}\,,\quad \phi_2=\pi+\frac{i}{2}\log\left(\frac{f}{\bar{f}}\right)\,,
\end{equation}
while the expressions for $\theta$ differ and are respectively given by
\begin{align}
\label{eq:real_uniton}
    \text{real uniton}:\qquad \sin^2\theta &= \frac{4|f|^2}{(1+|f|^2)^2+(\eta-\zeta)^2(1-|f|^2)^2}\,,\\
\label{eq:complex_uniton}
    \text{complex uniton}:\qquad \theta&=\frac{\pi}{2}+i\arctanh\left(\frac{1}{2}\left(|f|+\frac{1}{|f|}\right)\sqrt{(\eta-\zeta)^2+1} \right)\,.
\end{align}
The nomenclature `real' vs. `complex' is due to the fact that for the real uniton, $\theta$ is manifestly real and hence the group-valued field $g$ indeed lies in $\mathrm{SU}(2)$. In contrast, for the complex uniton $\theta$ is complex-valued and hence $g$ takes values in the complexified group $\mathrm{SL}(2,\mathbb{C})$.\footnote{Note that for real $x$
\begin{equation*}
    \mathrm{Im}\,\arctanh\,x=\begin{cases}
    0\,,& |x|<1\,,\\
    -\frac{\pi}{2}\mathrm{sign}(x)\,,& |x|>1\,.
    \end{cases}
\end{equation*}
Therefore, one finds that for the complex uniton $\mathrm{Re}\,\theta=\pi$, hence $g$ in fact takes values in $\mathrm{SU}(1,1)$. See also the discussion in section \ref{subsec:Weil_SL2}.}

For completeness, we also record the metric of the $\mathrm{SU}(2)$ bi-Yang--Baxter model when evaluated on the real and complex unitons. Introducing polar coordinates $f=r e^{i\alpha}$ and writing $R=r^2$ one finds\footnote{Here we have used the fact that
\begin{equation*}
    4R^2 \left(\frac{\mathrm{d}\theta}{\mathrm{d}R}\right)^2 = \sin^2\theta+(\zeta-\eta)^2\sin^4\theta\,,
\end{equation*}
which holds for both the real and complex uniton and can be verified by explicit computation.}
\begin{equation}
\label{eq:ds2_onshell}
    ds^2 = \frac{1}{\Delta}\left(\frac{\mathrm{d}\theta}{\mathrm{d}R}\right)^2\left(\mathrm{d}R^2+4R^2\mathrm{d}\alpha^2\right)\,,
\end{equation}
for both unitons. We see that in the target space the uniton solutions correspond to a squashed two-sphere inside $\mathrm{SU}(2)$. From here it follows from explicit integration that the unitons have finite action and we refer the reader to \cite{Schepers:2020ehn} for further details. One finds that the on-shell actions evaluate to (we neglect the overall factor coming from the angular integration)
\begin{equation}
    S_{\text{real uniton}} = \frac{1}{2\eta\zeta}\left[(\eta+\zeta)\arctan(\eta+\zeta)-(\eta-\zeta)\arctan(\eta-\zeta) \right]\,,
\end{equation}
and
\begin{equation}
    S_{\text{complex uniton}} = \frac{1}{2\eta\zeta}\left[(\eta+\zeta)\arccot(\eta+\zeta)-(\eta-\zeta)\arccot(\eta-\zeta) \right]\,.
\end{equation}
Here we have assumed the domain of integration to be the entire complex plane. In other words, $R$ ranges from 0 to infinity. In the next section we will encounter a situation where $f(z)$ instead takes values in the unit disk, in which case $R\in[0,1]$. One can verify that this only changes the above results by a factor of 1/2.

\subsection{Weil Operator as a Complex Uniton}
\label{subsec:Weil_SL2}

The unitons discussed in the previous section are solutions to the $\mathrm{SU}(2)$ bi-Yang--Baxter model. In this section, we will instead be concerned with the $\mathrm{SL}(2,\mathbb{R})$ bi-Yang--Baxter model. This model has a solution which naturally arises from the study of variations of Hodge structure, applied to the simplest example of a torus. More precisely, the solution is given by the so-called Weil operator which, roughly speaking, corresponds to the Hodge star when viewed as an operator on the middle de Rham cohomology of the torus. The Weil operator is a function of the Teichm\"uller parameter $\tau$ of the torus. This parameter is then reinterpreted as the worldsheet coordinate in the bi-Yang--Baxter model. Interestingly, this solution also satisfies $g^2=-1$, which suggests that it might be related to the uniton solutions. Indeed, we show that the Weil operator can be mapped to the \textit{complex} uniton, for a particular choice of holomorphic function $f(z)$, via a Cayley transformation.

\subsubsection*{Weil Operator}
Let us start by introducing the Weil operator of the torus. A convenient description of the torus $\mathbb{T}$ is as a lattice 
\begin{equation}
    \mathbb{T} = \mathbb{C}/(\mathbb{Z}+\tau\mathbb{Z})\,,\qquad \mathrm{Im}\,\tau>0\,,
\end{equation}
where $\tau$ is the Teichm\"uller parameter taking values in the complex (strict) upper half-plane. We parametrize the torus by two periodic coordinates $\xi_1,\xi_2$ with $\xi_i\sim\xi_i+1$. Then the metric on the torus can be written as
\begin{equation}
\label{eq:metric_torus}
    ds^2 = \frac{|\mathrm{d}\xi_1+\tau\,\mathrm{d}\xi_2|^2}{\mathrm{Im}\,\tau}\,.
\end{equation}
Here we have normalized the metric so that the torus has unit volume. The Weil operator is closely related to the Hodge star operator on the torus. The action of the Hodge star on the one-forms $\mathrm{d}\xi_1$ and $\mathrm{d}\xi_2$ follows directly from the metric \eqref{eq:metric_torus} and is given by
\begin{equation}
\label{eq:Hodge_star_torus}
    \star\,\mathrm{d}\xi_1 =\frac{\mathrm{Re}\,\tau}{\mathrm{Im}\,\tau}\mathrm{d}\xi_1+\frac{|\tau|^2}{\mathrm{Im}\,\tau}\mathrm{d}\xi_2\,,\qquad \star\,\mathrm{d}\xi_2 = -\frac{1}{\mathrm{Im}\,\tau}\mathrm{d}\xi_1-\frac{\mathrm{Re}\,\tau}{\mathrm{Im}\,\tau}\mathrm{d}\xi_2\,.
\end{equation}
To obtain the Weil operator, one should view this as an action on the middle de Rham cohomology of the torus. Indeed, in the basis $\{[\mathrm{d}\xi_1],[\mathrm{d}\xi_2]\}$, where $[\omega]$ denotes the equivalence class of a one-form $\omega$, the action of the Hodge star can be represented as a matrix
\begin{equation}
\label{eq:Weil_torus}
    C(x,y) = \frac{1}{y}\begin{pmatrix}
    x & -1\\
    x^2+y^2 & -x
    \end{pmatrix}\,,
\end{equation}
where we have set $\tau=x+iy$. We will refer to \eqref{eq:Weil_torus} as the Weil operator of the torus. Note that it is an element of $\mathrm{SL}(2,\mathbb{R})$.

\subsubsection*{Relation to the Complex Uniton}
An interesting property of \eqref{eq:Weil_torus} is that it satisfies\footnote{More generally, the origin of this relation is the fact that $\star\star$ evaluates to $\pm 1$, with the sign determined by the degree of the differential form it acts on and the dimension of the spacetime in question.}
\begin{equation}
    C^2=-1\,,
\end{equation}
which is shared by the unitons solutions discussed in section \ref{sec:bi_YB_unitons}. In fact, we will now argue that the Weil operator can be viewed as a complex uniton for a specific choice of the holomorphic function $f(z)$. 

As a preliminary remark, we stress that one cannot simply compare the expressions \eqref{eq:g_SU2} and \eqref{eq:Weil_torus}, as the two lie in different groups, namely $\mathrm{SU}(2)$ and $\mathrm{SL}(2,\mathbb{R})$, respectively. There is, however, a natural two-step procedure to pass between the two groups by combining a so-called Cayley transformation with an analytic continuation. Let us first elaborate on the former. We introduce the matrix
\begin{equation}
\label{eq:Cayley_sl2}
    \rho = \frac{1}{\sqrt{2}}\begin{pmatrix}
    1 & i\\
    i & 1
    \end{pmatrix}\,,
\end{equation}
which is an element of $\mathrm{SL}(2,\mathbb{C})$. Then it is straightforward to show that the adjoint action
\begin{equation}
    \mathrm{Ad}_\rho:\mathrm{SL}(2,\mathbb{R})\rightarrow\mathrm{SU}(1,1)
\end{equation}
is an isomorphism of real Lie groups, which is commonly referred to as a Cayley transformation. Indeed, it provides an interpolation between the two real forms $\mathrm{SL}(2,\mathbb{R})$ and $\mathrm{SU}(1,1)$ of $\mathrm{SL}(2,\mathbb{C})$. For the second step of the proposed procedure, one interpolates between $\mathrm{SU}(1,1)$ and $\mathrm{SU}(2)$ via an analytic continuation. In the parametrization \eqref{eq:g_SU2} this is straightforwardly given by sending $\theta\mapsto -i\theta$.

We now apply the above procedure to compare the Weil operator of the torus to a generic element in $\mathrm{SU}(2)$. Practically, it is easiest to take the expression in \eqref{eq:g_SU2}, analytically continue it to $\mathrm{SU}(1,1)$ by setting $\theta= i\tilde{\theta}$ and then apply $\mathrm{Ad}_{\rho^{-1}}$ to end up in $\mathrm{SL}(2,\mathbb{R})$. The result of this computation is
\begin{equation}
\label{eq:g_Sl2}
    g_{\mathrm{SL}(2,\mathbb{R})} = \begin{pmatrix}
    \cosh\tilde{\theta} \cos\phi_1 - \sinh\tilde{\theta} \sin\phi_2 & -\cosh\tilde{\theta}\sin\phi_1-\sinh\tilde{\theta}\cos\phi_2\\
    \cosh\tilde{\theta}\sin\phi_1-\sinh\tilde{\theta}\cos\phi_2 & \cosh\tilde{\theta}\cos\phi_1+\sinh\tilde{\theta}\sin\phi_2
    \end{pmatrix}\,,
\end{equation}
which is indeed an element of $\mathrm{SL}(2,\mathbb{R})$. Comparing \eqref{eq:Weil_torus} and \eqref{eq:g_Sl2} and solving for $\tilde{\theta},\phi_1,\phi_2$ in terms of $x,y$ gives
\begin{equation}
\label{eq:Weil_sol}
    \phi_1 = \frac{\pi}{2}\,,\qquad \phi_2 = \pi+\frac{i}{2}\log \left(\frac{f}{\Bar{f}}\right)\,,\qquad \tilde{\theta} = \frac{i\pi}{2}+\arctanh\left[\frac{1}{2}\left(|f|+\frac{1}{|f|}\right)\right]\,,
\end{equation}
where $f(z)$ is the following holomorphic function of the complexified worldsheet coordinates
\begin{equation}
\label{eq:f_mobius}
    f(z) = \frac{z-i}{z+i}\,,\qquad z=x+iy\,.
\end{equation}
Indeed, identifying the Teichm\"uller parameter $\tau$ with the worldsheet coordinate $z$ and recalling that $\theta=i\tilde{\theta}$, one sees that this solution is precisely of the form of a complex uniton \eqref{eq:complex_uniton} with additionally $\zeta=\eta$.\footnote{Strictly speaking, an exact match is obtained after sending $\theta\mapsto\theta+\pi$, corresponding to $C\mapsto -C$. Of course, the overall sign of $C$ is simply a convention.} The function $f(z)$ is a special type of M\"obius transformation that conformally maps the upper half-plane to the unit disc. Note that it is holomorphic on the upper half-plane, but has a first order pole at $z=-i$. 

\begin{figure}
     \centering
     \begin{subfigure}[b]{0.3\textwidth}
         \centering
         \includegraphics[width=\textwidth]{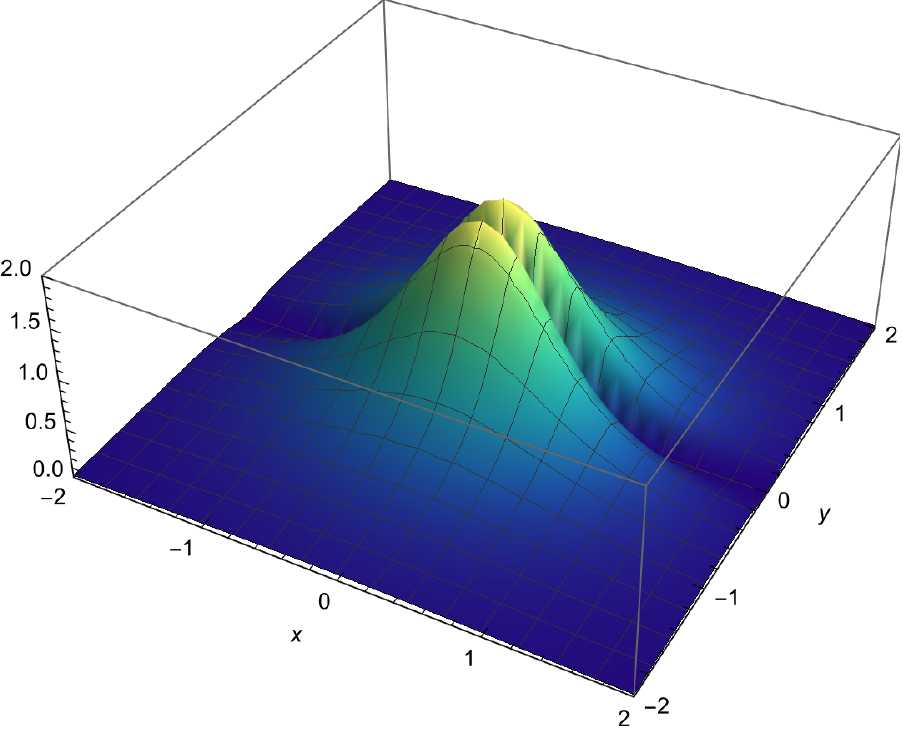}
         \caption{$\eta=0.7$, $\zeta=0.6$}
         \label{fig:plot1}
     \end{subfigure}
     \hfill
     \begin{subfigure}[b]{0.3\textwidth}
         \centering
         \includegraphics[width=\textwidth]{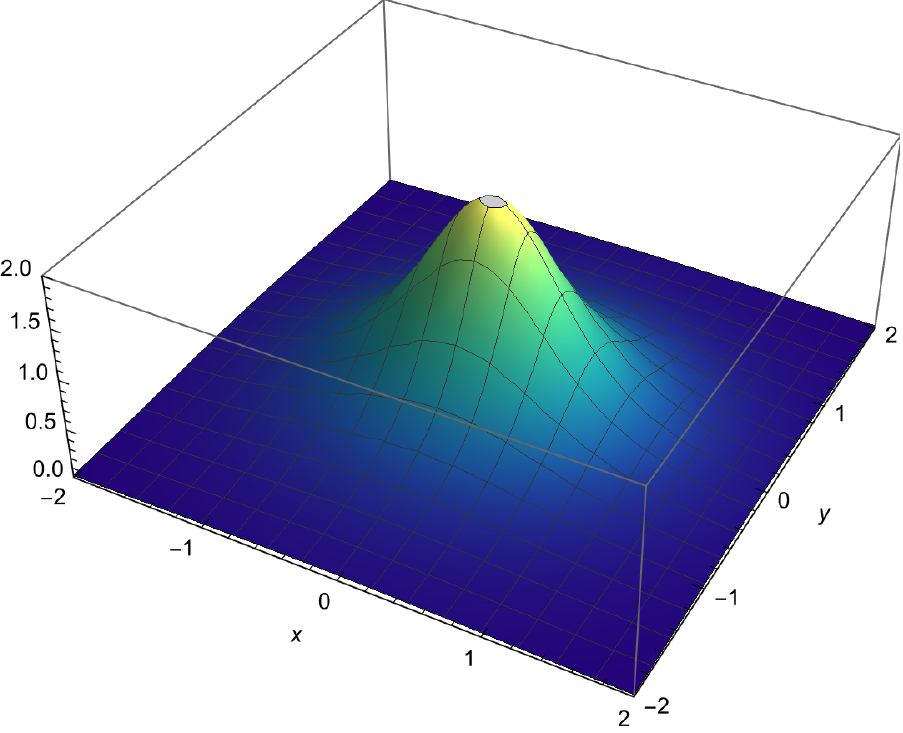}
         \caption{$\eta=0.7$, $\zeta=0.7$}
         \label{fig:plot2}
     \end{subfigure}
     \hfill
     \begin{subfigure}[b]{0.3\textwidth}
         \centering
         \includegraphics[width=\textwidth]{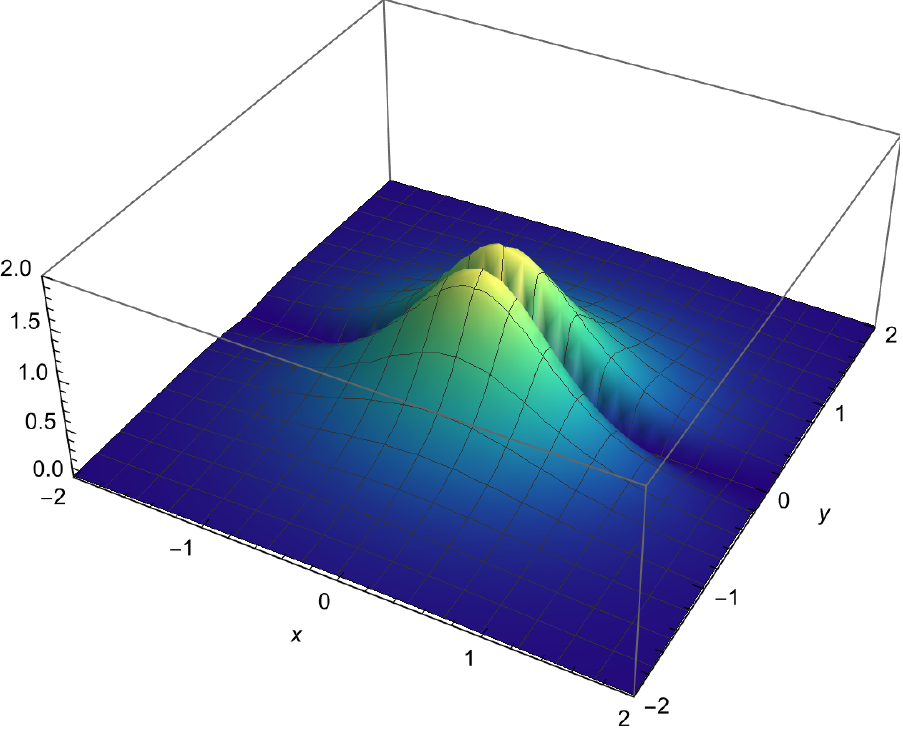}
         \caption{$\eta=0.7$, $\zeta=0.8$}
         \label{fig:plot3}
     \end{subfigure}
        \caption{Plot of the Lagrangian density of the $\mathrm{SU}(2)$ bi-Yang--Baxter model evaluated on the complex uniton solution with $f(z) = \frac{z-i}{z+i}$, for three values of the deformations parameters $(\eta,\zeta)$. For $\eta\neq\zeta$ there are two bumps, which coalesce on the critical line $\eta=\zeta$.}
        \label{fig:unitons}
\end{figure}

In figure \ref{fig:unitons} we have illustrated the Lagrangian density for the complex uniton defined by the particular holomorphic function \eqref{eq:f_mobius}. It is interesting to contrast this with the plots in \cite{Schepers:2020ehn}, where instead a linear function $f(z)= \frac{z}{2}$ was used. In both cases there is a clear transition at the critical point $\zeta=\eta$. On the other hand, the concentric valley structure in \cite{Schepers:2020ehn} is not present here. Rather, in our case the density is not rotationally invariant, but only invariant under reflections $x\mapsto -x$ and $y\mapsto -y$. Of course, this can be explained by noting that the Lagrangian density is a function of $|f(z)|$. 

It is also interesting to recall the $\mathbb{Z}_2$-symmetry \eqref{eq_Z2_cocritical} which maps the critical line to the co-critical line. Indeed, one finds that it acts on the Weil operator as
\begin{equation}
    C(x,y)\mapsto \frac{i}{y}\begin{pmatrix}
    -1 & -x\\
    x & x^2+y^2
    \end{pmatrix} = C(x,y)\cdot\begin{pmatrix}
    0 & i\\
    i & 0
    \end{pmatrix}\,.
\end{equation}
Interestingly, the $\mathbb{Z}_2$ transformation can be described as a right-multiplication of $C(x,y)$ by an element in $\mathrm{SL}(2,\mathbb{C})$ (even in $\mathrm{SU}(2)$). However, as a result the transformed Weil operator is no longer real-valued. Therefore, it will define a solution to the co-critical $\mathrm{SL}(2,\mathbb{C})$ bi-Yang--Baxter model, as can be verified by explicit computation.

As a final comment, one can use the result \eqref{eq:ds2_onshell}, taking $f(z)$ as above, to find that the on-shell metric is given by
\begin{equation}
    ds^2 = \frac{4|\mathrm{d}z|^2}{(z-\bar{z})^2-4\eta^2(1+|z|^2)^2}\,.
\end{equation}
Note that in the limit $\eta\rightarrow 0$ one recovers the standard metric on the Poincar\'e upper half-plane. 

\subsubsection*{R-matrix revisited}
As a result of the identification between the Weil operator and a complex uniton, one expects that the Weil operator provides a solution to the $\mathrm{SL}(2,\mathbb{R})$ bi-Yang--Baxter model. To make this precise, one should also identify the appropriate $R$-matrix by taking the $\mathrm{SU}(2)$ $R$-matrix in \eqref{eq:R_su2} and translating it to an endomorphism of $\mathrm{SL}(2,\mathbb{R})$ via the Cayley transform. Explicitly, this gives\footnote{One can check that the $R$-matrix remains unchanged under the analytic continuation from $\mathrm{SU}(2)$ to $\mathrm{SU}(1,1)$.}
\begin{equation}
    R_{\mathrm{SL}(2,\mathbb{R})} = \mathrm{Ad}_{\rho^{-1}}\circ R_{\mathrm{SU}(2)}\circ\mathrm{Ad}_\rho\,.
\end{equation}
One can verify that the resulting $R$-matrix acts as
\begin{equation}
\label{eq:R_Weil}
    R_{\mathrm{SL}(2,\mathbb{R})}\,H = -(E_++E_-)\,,\qquad R_{\mathrm{SL}(2,\mathbb{R})}\,E_\pm = \frac{1}{2}H\,,
\end{equation}
and is manifestly real. The reader is invited to check that indeed the Weil operator \eqref{eq:Weil_sol} is a solution to the critical $\mathrm{SL}(2,\mathbb{R})$ bi-Yang--Baxter model defined by the $R$-matrix \eqref{eq:R_Weil}. One may wonder how this $R$-matrix differs from the Drinfeld'd--Jimbo solution we started with. The answer is that the $R$-matrix \eqref{eq:R_Weil} is also a Drinfel'd--Jimbo solution, but for a different choice of Cartan generators. This will be explained further in section \ref{subsec:horizontal_sl2}.

\subsubsection*{An underlying $\mathrm{Sl}(2)$-orbit}
Let us return to the Weil operator \eqref{eq:Weil_torus} to elucidate a very particular underlying structure, which may not be immediately apparent from its matrix representation. Indeed, note that the Weil operator can be factorized in the following way
\begin{equation}
    C(x,y) = h(x,y)C_\infty h(x,y)^{-1}\,,
\end{equation}
where we have introduced
\begin{equation}
\label{eq:h_Cinfty}
    h(x,y) = \frac{1}{\sqrt{y}}\begin{pmatrix}
    1 & 0\\
    x & y
    \end{pmatrix}\,,\qquad C_\infty = \begin{pmatrix}
    0 & 1\\
    -1 & 0
    \end{pmatrix}\,.
\end{equation}
More abstractly, both $h(x,y)$ and $C_\infty$ can be written in terms of $\mathfrak{sl}(2,\mathbb{R})$-valued objects as
\begin{equation}
\label{eq:h_Cinfty_sl2}
    h(x,y) = e^{xN^-}y^{-\frac{1}{2}N^0}\,,\qquad C_\infty = (-1)^{Q_\infty}\,,
\end{equation}
where we have chosen to change our notation for the $\mathfrak{sl}(2,\mathbb{R})$ generators to
\begin{equation}
    N^\pm = E_\pm\,,\qquad N^0=H\,,\qquad 
\end{equation}
and introduced another operator
\begin{equation}
\label{eq:Q_sl2}
    Q_\infty = \frac{i}{2}\begin{pmatrix}
    0 & -1\\
    1 & 0
    \end{pmatrix}\,,
\end{equation}
for which $iQ_\infty$ is valued in $\mathfrak{sl}(2,\mathbb{R})$. At this point, there are a number of interesting observations we can make. First, the operator $h(x,y)$ in \eqref{eq:h_Cinfty} is of the form of a so-called $\mathrm{Sl}(2)$-orbit, which features prominently in the study of variations of Hodge structure, as will be explained further in section \ref{sec:Hodge_Theory}. Second, there is the appearance of the operator $Q_\infty$, which in previous works has been referred to as a \textit{charge operator}. One can verify that it satisfies the following commutation relations with the real $\mathfrak{sl}(2)$ triple $N^+,N^0,N^-$
\begin{equation}
\label{eq:Q_commutation_sl2}
    [Q_\infty, N^0] = i\left(N^+ + N^-\right)\,,\qquad [Q_\infty, N^\pm] = -\frac{i}{2}N^0\,.
\end{equation}
Such $\mathfrak{sl}(2)$-triples are referred to as \textit{horizontal} $\mathfrak{sl}(2)$-triples and we will discuss them in more generality in section \ref{sec:Hodge_Theory}. There is an intimate relationship between such horizontal $\mathfrak{sl}(2)$-triples and $\mathrm{Sl}(2)$-orbits, which will be elaborated upon in section \ref{subsec:asymp_Hodge_theory}.

As a final comment, one may compare \eqref{eq:Q_commutation_sl2} with the $R$-matrix \eqref{eq:R_Weil} to find that
\begin{equation}
    R_{\mathrm{SL}(2,\mathbb{R})}=\mathrm{ad}_{iQ_\infty}\,.
\end{equation}
In other words, we have found that the solution we have obtained for the critical $\mathrm{SL}(2,\mathbb{R})$ bi-Yang--Baxter model can be completely characterized by an operator $Q_\infty$ and an $\mathfrak{sl}(2,\mathbb{R})$-triple $N^+, N^0,N^-$ that is horizontal with respect to $Q_\infty$. It turns out that by studying the abstract properties of horizontal $\mathfrak{sl}(2)$-triples one can greatly generalize the simple solution we have considered here to groups of larger rank. This is the topic of the next section.

\section{Horizontal $\mathfrak{sl}(2)$-triples and $\mathrm{Sl}(2)$-orbits}
\label{sec:Hodge_Theory}

In the previous section we have argued that the Weil operator of the torus provides a solution to the critical $\mathrm{SL}(2,\mathbb{R})$ bi-Yang--Baxter model. In this section we show that this solution can be generalized to arbitrary groups as long as they admit at least one horizontal $\mathfrak{sl}(2)$-triple. In section \ref{subsec:horizontal_sl2} we define the latter in more detail and discuss how it defines a particular class of $R$-matrices. Given a horizontal $\mathfrak{sl}(2)$-triple we then write down the generalized solution in section \ref{subsec:sl2_orbits} and show explicitly that it solves the equations of motion of the bi-Yang--Baxter model associated to the mentioned class of $R$-matrices. The solutions we consider originate from the study of variations of Hodge structure and correspond to the so-called $\mathrm{Sl}(2)$-orbit approximation of the Weil operator. For the interested reader we have included in section \ref{subsec:asymp_Hodge_theory} a basic introduction to the field of asymptotic Hodge theory, where we discuss the Hodge-theoretic interpretation of the $\mathrm{Sl}(2)$-orbit and its relation to horizontal $\mathfrak{sl}(2)$-triples. Finally, in section \ref{subsec:example} we have included a very explicit example of a horizontal $\mathfrak{sl}(2)$-triple in $\mathfrak{sp}(4,\mathbb{R})$ and the associated $\mathrm{Sl}(2)$-orbit approximation of the Weil operator.  

\subsection{Horizontal $\mathfrak{sl}(2)$-triples}
\label{subsec:horizontal_sl2}

Let $\mathfrak{g}_{\mathbb{R}}$ be a real Lie algebra. Then a standard $\mathfrak{sl}(2)$-triple is a set of three elements in $\mathfrak{g}_{\mathbb{R}}$, denoted by $\{N^+,N^0,N^-\}$, satisfying the commutation relations
\begin{equation}
\label{eq:sl2_algebra}
    [N^0,N^\pm] =\pm 2 N^{\pm}\,,\qquad [N^+,N^-]=N^0\,.
\end{equation}
For classical Lie algebras $\mathfrak{g}_\mathbb{R}$, there is a classification of $\mathfrak{sl}(2)$-triples in $\mathfrak{g}_\mathbb{R}$ in terms of signed Young diagrams, see e.g.~\cite{McGovern:1993}. One can think of a horizontal $\mathfrak{sl}(2)$-triple as a standard triple with some additional properties which we will now discuss.

\subsubsection*{Charge Operator}
The additional properties that a horizontal $\mathfrak{sl}(2)$-triple must satisfy are most easily described in terms of an element $Q_{\infty}\in\mathfrak{g}_{\mathbb{C}}$ which will be referred to as a \textit{charge operator}. The charge operator should satisfy two important properties. First, it should be purely imaginary-valued, i.e.
\begin{equation}
    \bar{Q}_{\infty} = -Q_{\infty}\,,
\end{equation}
where the bar denotes complex conjugation. Second, $Q_{\infty}$ should be a so-called grading element. This means that the adjoint action $\mathrm{ad}\,Q_{\infty}$ has only integer eigenvalues. This definition of the charge operator has its origin in the study of Hodge structures. Indeed, as discussed in e.g.~\cite{Robles:2012}, one can think of a charge operator $Q_{\infty}$ as an infinitesimal Hodge structure. For the purpose of this section, however, it will not be necessary to employ this Hodge-theoretic interpretation. 

As a first example, one can verify that the $Q_\infty$ defined in \eqref{eq:Q_sl2} defines a charge operator in $\mathfrak{sl}(2,\mathbb{C})$. However, we stress that for a given Lie algebra $\mathfrak{g}_{\mathbb{R}}$ there can be many inequivalent charge operators. Indeed, taking the algebra $\mathfrak{sp}(4,\mathbb{R})$ as a more non-trivial example, one can verify that the following elements
\begin{equation}
    \frac{i}{2}\begin{pmatrix}
    0 & 0 & 0 & -3\\
    0 & 0 & -1 & 0\\
    0 & 1 & 0 & 0\\
    3 & 0 & 0 & 0
    \end{pmatrix}\,,\qquad
   \frac{i}{2}\begin{pmatrix}
   0 & -2 & -1 & 0\\
   2 & 0 & 0 & -1\\
   1 & 0 & 0 & -2\\
   0 & 1 & 2 & 0
   \end{pmatrix}\,,\qquad
   \frac{i}{2}\begin{pmatrix}
   0 & -3 & 0 & 0\\
   1 & 0 & -2 & 0\\
   0 & 2 & 0 & -1\\
   0 & 0 & 3 & 0
   \end{pmatrix}
\end{equation}
all define charge operators.\footnote{The symplectic pairing used in each of these cases differs, we refer the reader to \cite{Grimm:2021idu} for more details.} A general classification of horizontal $\mathfrak{sl}(2)$-triples was given by C.~Robles in \cite{Robles:2015}, see also \cite{Kerr2017}.

\subsubsection*{Horizontality}
Given a standard $\mathfrak{sl}(2)$-triple and a charge operator $Q_{\infty}$, we say that the triple is \textit{horizontal} (with respect to $Q_{\infty}$) if it satisfies the following commutation relations
\begin{equation}
\label{eq:Q_commutation}
    [Q_{\infty},N^0] = i\left(N^+ +N^-\right)\,,\qquad [Q_{\infty},N^{\pm}] = -\frac{i}{2}N^0\,.
\end{equation}
At first sight, these particular commutation relations are not very illuminating. One way to gain some insight on their meaning is to map the $\mathfrak{sl}(2)$-triple to $\mathfrak{su}(1,1)$ using the Cayley transform. In general, this can be written as
\begin{equation}
\label{eq:Cayley_general}
    \rho = e^{\frac{i\pi}{4}\left(N^++N^-\right)}\,,
\end{equation}
which reduces to \eqref{eq:Cayley_sl2} for the fundamental representation. Let us then introduce the complex triple
\begin{equation}
    L_{\alpha} = \rho N^\alpha\rho^{-1}\,,\qquad \alpha= 1,0,-1\,.
\end{equation}
By explicit computation using the commutation relations \eqref{eq:sl2_algebra} one finds
\begin{equation}
    L_{\pm 1} = \frac{1}{2}\left(N^+ + N^-\mp i N^0 \right)\,,\qquad L_0 = -i(N^+-N^-)\,.
\end{equation}
Furthermore, the relations \eqref{eq:Q_commutation} can be written in terms of the $L_\alpha$ as
\begin{equation}
\label{eq:Q_commutation_complex}
    [Q_{\infty},L_\alpha] = \alpha L_\alpha\,.
\end{equation}
Importantly, we see that $L_0$ commutes with $Q_{\infty}$. In fact, defining $\hat{Q}_{\infty} = Q_{\infty}-\frac{1}{2}L_0$, one immediately sees that $\hat{Q}_{\infty}$ commutes with all $L_\alpha$. One similarly sees that $\hat{Q}_\infty$ commutes with all the $N^\alpha$ and that $i\hat{Q}_\infty$ is real. Therefore, one can think of the existence of a horizontal $\mathfrak{sl}(2)$-triple as an underlying $\mathrm{SL}(2,\mathbb{R})\times\mathrm{U}(1)$ symmetry. For this reason we refer to $Q_{\infty}$ as a charge operator, as it can be interpreted as parametrizing an additional $\mathrm{U}(1)$ symmetry. The reverse statement, however, is not necessarily true. Indeed, having an additional $\mathrm{U}(1)$ symmetry does not guarantee that the corresponding charge operator is really a grading operator. Furthermore, there are examples of horizontal $\mathfrak{sl}(2)$-triples for which the operator $\hat{Q}_\infty$ vanishes. 

\subsubsection*{R-matrix from horizontal $\mathfrak{sl}(2)$-triples}

In the next section, we will argue that a horizontal $\mathfrak{sl}(2)$-triple provides a solution to the critical bi-Yang--Baxter model, for a particular class of $R$-matrices. These $R$-matrices must be compatible with the $\mathfrak{sl}(2)$-triple in the sense that they must act in the same way as we found in the simple example of section \ref{subsec:Weil_SL2}. Let us make this more precise.

Given a charge operator $Q_{\infty}$ and a horizontal $\mathfrak{sl}(2)$-triple in a Lie algebra $\mathfrak{g}$, we impose that the $R$-matrix acts as
\begin{equation}
\label{eq:def_R_general}
    R\,Q_{\infty}=0\,,\qquad R\,N^0 = -(N^++N^-)\,,\qquad R\,N^\pm = \frac{1}{2}N^0\,.
\end{equation}
One readily checks that the modified classical Yang--Baxter equation is then satisfied for this $\mathfrak{sl}(2)$-triple. In terms of the complex generators $L_\alpha$, it acts as
\begin{equation}
    R\,Q_{\infty}=0\,,\qquad R\,L_0 = 0\,,\qquad R\,L_{\pm 1} = \pm i L_{\pm 1}\,.
\end{equation}
Comparing this to \eqref{eq:def_R_DJ}, we see that this choice of $R$-matrix is nothing but the Drinfel'd--Jimbo solution, where we have chosen $L_0$ and $Q_{\infty}$ as the generators of a Cartan subalgebra of $\mathfrak{sl}(2,\mathbb{C})\times\mathfrak{u}(1)$ and $L_{\pm 1}$ correspond to the positive and negative roots.
However, in general this does not yet specify the full $R$-matrix, as its action on the remaining generators of $\mathfrak{g}$ is not yet fixed. Indeed, if the rank of $\mathfrak{g}$ is greater than two, one must identify additional Cartan generators beyond $L_0$ and $Q_{\infty}$ to complete the full Drinfel'd--Jimbo solution.

\subsection{$\mathrm{Sl}(2)$-orbits and generalized unitons}
\label{subsec:sl2_orbits}

In this section we will introduce a generalization of the Weil operator of the torus discussed in section \ref{subsec:Weil_SL2} using purely the data of a horizontal $\mathfrak{sl}(2)$-triple. Hodge-theoretically, the operators we discuss correspond to so-called $\mathrm{Sl}(2)$-orbits. This point of view is expanded upon in section \ref{subsec:asymp_Hodge_theory}. In the following, however, we will purely use the abstract properties of the horizontal $\mathfrak{sl}(2)$-triple to show that such operators also solve the critical bi-Yang--Baxter model and therefore provide a generalization of the complex unitons to groups beyond $\mathrm{SU}(2)$. 

\subsubsection*{$\mathrm{Sl}(2)$-orbits}

In section \ref{subsec:Weil_SL2} we highlighted a particular underlying structure in the Weil operator of the torus in terms of two elements $h(x,y)$ and $C_\infty$, see \eqref{eq:h_Cinfty} and \eqref{eq:h_Cinfty_sl2}. The idea is simply to generalize this to arbitrary horizontal $\mathfrak{sl}(2)$-triples. In other words, we introduce
\begin{equation}
\label{eq:def_h}
    h(x,y) = e^{xN^-}y^{-\frac{1}{2}N^0}\,,\qquad C_\infty = (-1)^{Q_\infty}
\end{equation}
and define
\begin{equation}
\label{eq:def_Weil_general}
    g(x,y) = h(-1)^{Q_\infty}h^{-1}\,.
\end{equation}
We will refer to \eqref{eq:def_Weil_general} as simply the Weil operator. To be precise, one should refer to $g(x,y)$ as the $\mathrm{Sl}(2)$-orbit of a boundary Weil operator, see section \ref{subsec:asymp_Hodge_theory}. Similarly $h(x,y)$ should be referred to as the $\mathrm{Sl}(2)$-orbit approximation of the period map. However, we will loosely refer to all these objects as $\mathrm{Sl}(2)$-orbits. Note that $g(x,y)$ satisfies a `twisted boundary condition' 
\begin{equation}
    g(x+1,y) = e^{N^-} g(x,y) e^{-N^-}\,,
\end{equation}
which is similar to the adiabatic reduction used in \cite{Schepers:2020ehn} to reduce the $\mathrm{SU}(2)$ bi-Yang--Baxter model to a model of quantum mechanics. 

\subsubsection{Generalized Unitons}

We now explicitly show that the Weil operator $g$ in \eqref{eq:def_Weil_general}, together with the $R$-matrix defined in \eqref{eq:def_R_general} solves the equations of motion of the critical bi-Yang--Baxter model. Here we will focus on the main steps in the calculation, and refer the reader to appendix \ref{app:computations} where further details are given. More precisely, throughout the calculation it is necessary to know how $\mathrm{Ad}_h$, $\mathrm{Ad}_{h^{-1}}$ and $\mathrm{Ad}_g$ act on the $\mathfrak{sl}(2,\mathbb{R})$-triple. This can be straightforwardly computed using the commutation relations, and the results are listed in equations \eqref{eq:Adh_N+}-\eqref{eq:Adhinv_N-} and \eqref{eq:Adg_N+}-\eqref{eq:Adg_N-}.

To start, it will be most convenient to present the equations of motion in real coordinates $x,y$ in terms of which they read
\begin{equation}
    \partial_x J_y - \partial_y J_x - \eta[J_x,J_y]=0\,,
\end{equation}
where we recall the $R$-bracket in \eqref{eq:R_bracket} and denote
\begin{equation}
    J_x = J_+ + J_-\,,\qquad J_y = -i\left(J_+ - J_-\right)\,,
\end{equation}
with $J_\pm$ defined in \eqref{eq:def_Jpm}. For clarity of presentation, we will divide the main computation into three steps.

\subsubsection*{Step 1: $j_\mu$}
As a first step, let us compute the simpler objects
\begin{equation}
    j_\mu = g^{-1}\partial_\mu g\,.
\end{equation} 
Using \eqref{eq:def_Weil_general} one has
\begin{equation}
    j_\mu = h\left[(-1)^{\mathrm{ad}\,Q_\infty}h^{-1}\partial_\mu h - h^{-1}\partial_\mu h\right]h^{-1}\,.
\end{equation}
Then, using the expression \eqref{eq:def_h} for $h$, one readily computes
\begin{equation}
    h^{-1}\partial_x h = \frac{1}{y}N^-\,,\qquad h^{-1}\partial_y h = -\frac{1}{2y}N^0\,.
\end{equation}
Therefore, to compute $j_\mu$ it remains to apply the commutation relations \eqref{eq:Q_commutation} and evaluate the action of $\mathrm{Ad}_h$. For the first step we use the relations \eqref{eq:dagger_N+}-\eqref{eq:dagger_N-} and for the second step we use \eqref{eq:Adh_N+}-\eqref{eq:Adh_N-}. The result of this computation is
\begin{align}
\label{eq:jx}
    j_x &= -\frac{1}{y^2}\left[N^+-x N^0-(x^2-y^2)N^- \right]\,,\\
\label{eq:jy}
    j_y &= \frac{1}{y}\left[N^0+2x N^-\right]\,.
\end{align}

\subsubsection*{Step 2: $J_\mu$}
To proceed, we would like to compute the currents $J_\mu$, for which we must know the action of $R$ and $R_g$ on the currents $j_\mu$. The general action of these two operators on the $\mathfrak{sl}(2)$-triple follows from the definition \eqref{eq:def_R_general} and the action of $\mathrm{Ad}_g$ on the $\mathfrak{sl}(2)$-triple, which we record in \eqref{eq:Adg_N+}-\eqref{eq:Adg_N-}. Applying this to $j_x$ and $j_y$ gives
\begin{align}
    \left(\eta R+\zeta R^g\right)j_x &= \eta \frac{\alpha}{y}\left[N^0+2x N^-\right]+\frac{\zeta-\eta}{y^2}\left[xN^++\left(\frac{\alpha}{2}-x^2\right)N^0+x(1+2\alpha)N^-\right]\,,\\
     \left(\eta R+\zeta R^g\right)j_y&=\eta\frac{\alpha}{y^2}\left[N^+-x N^0-(x^2-y^2)N^-\right]\\
     &\qquad +\frac{\zeta-\eta}{y^3}\left[-(1+x^2)N^++x(1+x^2)N^0+(x^2+x^4-y^4)N^-\right]\,.\nonumber
\end{align}
where we have introduced
\begin{equation}
    \alpha=\alpha(x,y) = \frac{1+x^2+y^2}{y}\,.
\end{equation}
Crucially, one sees that the result simplifies considerably when evaluated on the critical line $\zeta=\eta$ and becomes
\begin{align}
    \left(R+R^g\right)j_x &= \frac{\alpha(x,y)}{y}\left[N^0+2x N^-\right]\,,\\
    \left(R+R^g\right)j_y &= \frac{\alpha(x,y)}{y^2}\left[N^+-x N^0-(x^2-y^2)N^-\right]\,,
\end{align}
The main observation is now the following. Comparing the above result with the expressions \eqref{eq:jx}-\eqref{eq:jy} for $j_x$ and $j_y$, we see that
\begin{equation}
    (R+R^g)j_x = \alpha(x,y)\,j_y\,,\qquad (R+R^g)j_y = -\alpha(x,y)\,j_x\,.
\end{equation}
In other words, for this particular choice of $g$ the currents $j_+$ and $j_-$ are eigenvectors of $R+R_g$. As a result, computing $J_x$ and $J_y$ is straightforward and gives
\begin{equation}
\label{eq:Jmu_result}
    J_x = \beta(x,y)j_y\,,\qquad J_y = -\beta(x,y)j_x\,,
\end{equation}
where we have introduced the function
\begin{equation}
\label{eq:def_beta}
    \beta(x,y) = \frac{1}{i-\alpha(x,y)\eta}\,.
\end{equation}

\subsubsection*{Step 3: equations of motion}
For the final step, we show that \eqref{eq:Jmu_result} solves the equations of motion
\begin{equation}
    \partial_x J_y - \partial_y J_x - \eta[J_x,J_y]=0\,.
\end{equation}
Indeed, inserting \eqref{eq:Jmu_result} one effectively needs to show that
\begin{equation}
    \beta\left(\partial_x j_x+\partial_y j_y\right) + \left(\partial_x\beta j_x+\partial_y\beta j_y + \eta [j_x,j_y]_R\right) = 0\,.
\end{equation}
This follows from straightforwardly using the definition of $\beta(x,y)$, inserting our results for $j_x$ and $j_y$, see \eqref{eq:jx}-\eqref{eq:jy}, and evaluating the $R$-bracket using \eqref{eq:R_bracket} and \eqref{eq:def_R_general}. In fact, the two terms in brackets vanish separately. Note that for the first term this is simply the statement that the one-form $\star j$ is exact, i.e.
\begin{equation}
\label{eq:star_j}
    \mathrm{d}\,\star\,j=0\,.
\end{equation}
Stated differently, $g$ also solves the principal chiral model. This is of course not surprising if one expects the solution to hold for all values of $\eta$, since in the limit $\eta\rightarrow 0$ the critical bi-Yang--Baxter model reduces precisely to the principal chiral model.

\subsubsection{Discussion}

To summarize, we have shown that for any real Lie group $G$ whose Lie algebra $\mathfrak{g}$ contains a horizontal $\mathfrak{sl}(2)$-triple, the Weil operator defined in \eqref{eq:def_Weil_general} provides a solution to the critical bi-Yang--Baxter model associated to the class of $R$-matrices satisfying \eqref{eq:def_R_general}. As this constitutes the main result of this work, let us make some additional comments.  

\subsubsection*{Finite Action}
As a first remark, we note that it is straightforward to show that our solutions have finite action. Indeed, one finds that
\begin{equation}
    S_{\mathrm{on-shell}} \sim \int d^2\sigma\,\beta \left(j_+,j_-\right)\,,
\end{equation}
where we recall that $\beta$ is defined by \eqref{eq:def_beta}. We stress that our solution is completely defined in terms of the horizontal $\mathfrak{sl}(2)$-triple and the pairing $(j_+,j_-)$ is independent of the particular representation of the triple, except for possibly an overall coordinate-independent factor. Therefore, it suffices to compute the on-shell action for the fundamental representation, where it was shown that the result is finite, as discussed in section \ref{subsec:unitons}. 

\subsubsection*{Multiple $\mathfrak{sl}(2)$-triples}

As a second remark, we note that it is entirely possible that $\mathfrak{g}$ contains multiple inequivalent (commuting) $\mathfrak{sl}(2)$-triples that are all horizontal with respect to the same charge operator. Of course, one can construct a corresponding Weil operator as in \eqref{eq:def_Weil_general} for each of these triples and obtain multiple inequivalent solutions to the corresponding bi-Yang--Baxter models. Interestingly, in Hodge theory the presence of multiple $\mathfrak{sl}(2)$-triples gives rise to so-called multi-variable $\mathrm{Sl}(2)$-orbits, which take the similar form
\begin{equation}
    g = h(-1)^{Q_\infty}h^{-1}\,,
\end{equation}
but with $h$ given by
\begin{equation}
    h(x_1,\ldots, x_n, y_1,\ldots, y_n) = \prod_{i=1}^n h_i(x_i, y_i)\,,\qquad h_i(x_i,y_i) = e^{x_i N^-_i}y_i^{-\frac{1}{2}N^0_i}\,,
\end{equation}
where $i$ enumerates the various horizontal $\mathfrak{sl}(2)$-triples. Of course, to view this as a solution of the bi-Yang--Baxter models one has to view one of the coordinates $t_i=x_i+iy_i$, for a fixed $i$, as the worldsheet coordinate and the others as some additional parameters. In this case one readily sees that this also defines a solution by exactly the same arguments above, as the transformation
\begin{equation}
    g\mapsto a\cdot g\cdot a^{-1}\,,\qquad R\mapsto \mathrm{Ad}_a\circ R\circ \mathrm{Ad}_a^{-1}\,,\qquad a\in G\,,
\end{equation}
is a global symmetry of the bi-Yang--Baxter model. It would, however, also be interesting to see if there exists a natural extension of the model to higher dimensions for which the multi-variable $\mathrm{Sl}(2)$-orbits provide a solution. 

\subsubsection*{Generalizations and relations to other integrable models}

As a third remark, we would like to stress that we have shown that the Weil operator solves the \textit{critical} bi-Yang--Baxter model, i.e. when $\zeta=\eta$. Additionally, it is straightforward to check that it does not solve the non-critical model, as we showed explicitly in section \ref{subsec:Weil_SL2} for the Weil operator of the torus. However, it is possible that by a suitable generalization of the ansatz one can also find solutions of the non-critical model. For example, for the $\mathrm{SU}(2)$ model one can apply the $\mathbb{Z}_2$-symmetry \eqref{eq_Z2_cocritical} to obtain instead a solution of the co-critical model and the same can be done for the $\mathrm{SL}(2,\mathbb{R})$ model. It would therefore be interesting to see if this symmetry, or an appropriate generalization thereof, also applies for the bi-Yang--Baxter model based on other groups.

Another point that is worth emphasizing is the fact that the critical bi-Yang--Baxter model is related to other integrable models. A trivial example is the limit $\eta\rightarrow 0$, for which it reduces to the principal chiral model. As mentioned earlier, since our solution holds for any value of $\eta$, this implies that the Weil operator also solves the principal chiral model, for which the equations of motion are simply the harmonicity condition \eqref{eq:star_j}. Relations to other integrable models have been established as well in the literature. For example, as mentioned before, the critical $\mathrm{SU}(2)$ bi-Yang--Baxter model is equivalent to the coset $\mathrm{SO}(4)/\mathrm{SO}(3)$ $\eta$-model \cite{Delduc:2013qra,Hoare:2014oua}. Furthermore, at the conformal point $\eta=\frac{i}{2}$ it coincides with the $\mathrm{SU}(1,1)/\mathrm{U}(1)$ gauged WZW model with an additional $\mathrm{U}(1)$ boson \cite{Hoare:2014oua}. Therefore the Weil operator (of the torus) also provides a solution to these models. There has also been work on relating the bi-Yang--Baxter model to generalized $\lambda$-deformations via Poisson-Lie T-duality. This was first worked out for the $\mathrm{SU}(2)$ model in \cite{Sfetsos:2015nya} and later proved in general in \cite{Klimcik:2016rov}, see also \cite{Klimcik:2015gba,Vicedo:2015pna,Hoare:2015gda}. Therefore, via this duality it is reasonable to expect that our solution can be mapped to solutions of generalized $\lambda$-deformations. This points to the striking idea that Hodge theory allows one to construct solutions to many integrable models. 

Indeed, as a last comment let us compare the results obtained in this work with our previous work \cite{Grimm:2021idu}. There it was shown that the Weil operator of an arbitrary variation of Hodge structure solves the $\lambda$-deformed $G/G$ model when $|\lambda|=1$. Indeed, in \cite{Grimm:2021idu} we considered the ansatz
\begin{equation}
\label{eq:g_previous}
    g = h\,z^{Q_\infty}h^{-1}\,,
\end{equation}
where $h$ reduces to \eqref{eq:def_h} in the $\mathrm{Sl}(2)$-orbit approximation. In that case, when also $z=-1$, the expression \eqref{eq:g_previous} coincides with \eqref{eq:def_Weil_general}. Furthermore, one can easily check that what was referred to as the `horizontality condition' in \cite{Grimm:2021idu} reduces to the condition that the $\mathfrak{sl}(2)$-triple is horizontal with respect to $Q_\infty$. It would be interesting to investigate how the relation among the solutions translates into a relation of the underlying integrable models. 

\subsection{Hodge-theoretic origin of $\mathrm{Sl}(2)$-orbits}
\label{subsec:asymp_Hodge_theory}

In this section we provide a basic introduction to the field of asymptotic Hodge theory and the origin of $\mathrm{Sl}(2)$-orbits. Let us first sketch the main ideas before we go into the details. The main idea is to generalize some of the structures that appeared in our discussion of the torus in section \ref{subsec:Weil_SL2}. The central ingredient is a number of vector spaces varying over some base. In the example of the torus, the vector spaces are the cohomology groups and the variation is parametrized by the Teichm\"uller parameter $\tau$. More generally, the vector spaces need to define a Hodge structure. If furthermore the variation of this Hodge structure over the base satisfies a particular criterion, then it is referred to as a variation of Hodge structure. In asymptotic Hodge theory, one is in particular interested in the general features of such variations in asymptotic regions of the parameter space, e.g.~when considering the limit $\tau\rightarrow i\infty$ for the torus. The striking fact is that the asymptotic behaviour of a variation of Hodge structure is universal, and (in a precise sense) asymptotes to an $\mathrm{Sl}(2)$-orbit. 

\subsubsection{The Torus Revisited}

As a warm-up, let us revisit the example of the torus. First, recall the decomposition of the middle de Rham cohomology
\begin{equation}
\label{eq:HS_torus}
    H^1(\mathbb{T},\mathbb{C}) = H^{1,0}\oplus H^{0,1}\,,\qquad H^{0,1} = \overline{H^{1,0}}\,,
\end{equation}
into holomorphic and anti-holomorphic harmonic one-forms. The decomposition \eqref{eq:HS_torus} is referred to as a Hodge decomposition. In the basis of $\{[\mathrm{d}\xi_1],[\mathrm{d}\xi_2]\}$ the subspaces can be represented as
\begin{equation}
\label{eq:Hpq_torus}
    H^{1,0}=\mathrm{span}_{\mathbb{C}}\begin{pmatrix}
    1\\\tau
    \end{pmatrix}\,,\qquad H^{0,1}= \mathrm{span}_{\mathbb{C}}\begin{pmatrix}
    1\\ \bar{\tau}
    \end{pmatrix}\,.
\end{equation}
In particular, while the full vector space $H^1$ is of course $\tau$-independent, the two subspaces $H^{1,0}$ and $H^{0,1}$ do depend on $\tau$. In fact, the dependence is such that \eqref{eq:HS_torus} defines a variation of Hodge structure. This will be properly defined shortly. 

\subsubsection*{Hodge Filtration}
However, to do so it will be convenient to introduce a slight reformulation. Indeed, consider the following two vector spaces
\begin{equation}
    F^1 = H^{1,0}\,,\qquad F^0 = H^{1,0}\oplus H^{0,1}\,,
\end{equation}
which clearly define a decreasing filtration 
\begin{equation}
    F^1\subseteq F^0\,.
\end{equation}
The spaces $F^p$ are collectively referred to as a Hodge filtration. They are an equivalent way of packaging the data of a Hodge decomposition. Indeed, one can recover the decomposition \eqref{eq:Hpq_torus} by
\begin{equation}
\label{eq:HS_HF_relation}
    H^{p,q} = F^p\cap \bar{F}^q\,.
\end{equation}
For this particular example, one easily verifies the following relations
\begin{equation}
\label{eq:horizontality_torus}
    \frac{\partial}{\partial\tau}F^1 \subseteq F^{0}\,,\qquad \frac{\partial}{\partial\bar{\tau}}F^1 \subseteq F^1\,.
\end{equation}
Let us explain the notation here. $F^1$ is a vector space spanned by the vector $(1,\tau)$. This vector can be differentiated with respect to $\tau$ to yield a new vector $(0,1)$, which clearly does not lie in $F^1$ but does lie in $F^0$. Similarly, differentiating with respect to $\bar{\tau}$ yields $(0,0)$, which does lie in $F^1$ as it is the zero vector. The two relations \eqref{eq:horizontality_torus} together are referred to as the horizontality condition. It is exactly this condition that will later be taken as the definition of a variation of Hodge structure (with an appropriate generalization).

\subsubsection*{Asymptotic Behaviour}

Before moving on to such generalizations, let us consider the limit $\tau\rightarrow i\infty$. From the point of view of the Hodge decomposition \eqref{eq:HS_torus} it appears difficult to make sense of this limit, as the vectors in \eqref{eq:Hpq_torus} diverge. However, in terms of the Hodge filtration, one can make the following observation\footnote{The reason for the notation $F^p_0$ is because, heuristically, it can be viewed as evaluating the filtration $F^p$ at the point $\tau=0$.}
\begin{equation}
\label{eq:nilpotent_orbit_torus}
    F^p = e^{\tau N} F^p_0\,,
\end{equation}
where
\begin{equation}
    F^1_0 = \mathrm{span}_\mathbb{C} \begin{pmatrix}
    1 \\ 0
    \end{pmatrix}\,,\qquad F^0_0 = \mathrm{span}_\mathbb{C} \left\{\begin{pmatrix}
    1 \\ 0
    \end{pmatrix}, \begin{pmatrix}
    0 \\ 1
    \end{pmatrix}\right\}\,,
\end{equation}
and
\begin{equation}
\label{eq:monodromy_torus}
    N = \begin{pmatrix}
    0 & 0 \\
    1 & 0
    \end{pmatrix}\,.
\end{equation}
Note that $N$ is exactly the standard lowering operator in $\mathfrak{sl}(2,\mathbb{R})$. This is a particular feature of the torus, due to its simplicity. In more complicated examples $N$ will be related to a lowering operator, but not exactly equal to it. What is, however, a general feature is that the matrix $N$ is nilpotent. Indeed, in this case we have $N^2=0$.

The point of the rewriting \eqref{eq:nilpotent_orbit_torus} is that there is a very natural way to make sense of the limit $\tau\rightarrow i\infty$. Indeed, since all the $\tau$-dependence is captured by the exponent $e^{\tau N}$, one defines the limiting filtration by simply stripping off this factor
\begin{equation}
    \lim_{\tau\rightarrow i\infty} e^{-\tau N}F^p = F_0^p\,.
\end{equation}
There is, however, one issue that arises from this construction. Indeed, applying \eqref{eq:HS_HF_relation} to define a tentative limiting Hodge decomposition associated to $F^p_0$, one finds
\begin{equation}
    H^{1,0}_0 = H^{0,1}_0= \mathrm{span}_\mathbb{C}\begin{pmatrix}
    1\\0
    \end{pmatrix}\,.
\end{equation}
In particular, $H^{1,0}_0\oplus H^{0,1}_0$ no longer returns the full vector space $H^{1}$. In other words, just the vector spaces $F^p_0$ are not sufficient to define an actual Hodge structure associated to the limit $\tau\rightarrow i\infty$. The missing ingredient, as one may have expected, is the matrix $N$ in \eqref{eq:monodromy_torus}. Indeed, note that
\begin{equation}
    N\begin{pmatrix}
    1 \\0
    \end{pmatrix} = \begin{pmatrix}
    0\\1
    \end{pmatrix}\,,
\end{equation}
which is the piece which is not captured by $H_0^{1,0}$ and $H_0^{0,1}$. Therefore, it is natural to expect that one should combine $N$ and $F^p_0$ in some way to define an actual Hodge structure. It turns out that the following definition does the job
\begin{equation}
\label{eq:Finfty_torus}
    F_\infty^p = e^{i N}F^p_0\,.
\end{equation}
Indeed, applying \eqref{eq:HS_HF_relation} one finds
\begin{equation}
    H^{1,0}_\infty = \mathrm{span}_\mathbb{C} \begin{pmatrix}
    1\\ i
    \end{pmatrix}\,,\qquad H^{0,1}_\infty = \mathrm{span}_\mathbb{C} \begin{pmatrix}
    1\\ -i
    \end{pmatrix}\,,
\end{equation}
which does define a proper Hodge decomposition. Hence, we have successfully associated a Hodge structure to the limit $\tau\rightarrow i\infty$, for which the decomposition \eqref{eq:HS_torus} still holds. To connect a bit with our discussion in section \ref{subsec:Weil_SL2}, we note that one can also describe the spaces $H^{1,0}_\infty$ and $H^{0,1}_\infty$ as the eigenspaces of the operator
\begin{equation}
    Q_\infty = \frac{i}{2}\begin{pmatrix}
    0 & -1\\
    1 & 0
    \end{pmatrix}\,,
\end{equation}
which also appeared in \eqref{eq:Q_sl2}. To be precise, elements in $H^{1,0}_\infty$ and $H^{0,1}_\infty$ will have eigenvalues $\frac{1}{2}$ and $-\frac{1}{2}$ respectively. 

\subsubsection*{Appearance of the $\mathrm{Sl}(2)$-orbit and Weil operator}

As a final question, one may wonder if it is possible to recover the full variation of Hodge structure from the boundary Hodge structure. In other words, we search for an operator, denoted by $h$, such that
\begin{equation}
\label{eq:def_h_HS}
    H^{p,q} = h\cdot H^{p,q}_\infty\,,
\end{equation}
for all $p,q$. The map $h$ is referred to as the period map. In this particular example this is relatively straightforward to obtain, and the result is
\begin{equation}
\label{eq:hsl2_torus}
    h = e^{xN} y^{-\frac{1}{2}N^0}\,,\qquad \tau=x+iy\,,
\end{equation}
where again
\begin{equation}
    N^0 = \begin{pmatrix}
    1 & 0\\
    0 & -1
    \end{pmatrix}
\end{equation}
is the standard grading operator in $\mathfrak{sl}(2,\mathbb{R})$. Note that \eqref{eq:hsl2_torus} is exactly what was considered in section \ref{subsec:Weil_SL2}. This is where the name $\mathrm{Sl}(2)$-orbit comes from. Indeed, we have found that the variation of Hodge structure defined by the middle de Rham cohomology of the torus can be viewed as the orbit of the boundary Hodge structure associated to the limit $\tau\rightarrow i\infty$ under an element of $\mathrm{SL}(2,\mathbb{R})$. Note that this relation is exact. Strikingly, it turns out that the appearance of $\mathrm{Sl}(2)$-orbits is a universal feature of variations of Hodge structure. In a precise sense, they characterize the leading order behaviour of the Hodge decomposition in the asymptotic regions of the parameter space. 

To end this example, we also explain the role of the Weil operator. Given a Hodge structure $H^{p,q}$, the Weil operator $C$ associated to this Hodge structure is defined by
\begin{equation}
    Cv = i^{p-q}v\,,\qquad v\in H^{p,q}\,.
\end{equation}
Naturally, given the relation \eqref{eq:def_h_HS} one can write it as
\begin{equation}
    C = h C_\infty h^{-1}\,,
\end{equation}
where $C_\infty$ is the Weil operator associated to the boundary Hodge structure $H^{p,q}_\infty$. It is straightforward to show that $C_\infty = (-1)^{Q_\infty}$. 

\subsubsection{General Variations of Hodge Structure}

Let us now generalize the above discussion to more general variations of Hodge structure. For simplicity, we will restrict to one-parameter variations, and denote the parameter by $z=x+iy$. Let $H$ be a complex vector space. Then a weight $D$ Hodge structure can be described by a Hodge decomposition
\begin{equation}
    H = H^{D,0}\oplus\cdots \oplus H^{0,D} = \bigoplus_{p+q=D}H^{p,q}\,,\qquad H^{q,p}=\overline{H^{p,q}}\,,
\end{equation}
where each of the vector spaces $H^{p,q}$ depends on $z$. As a motivating example, one may consider the primitive middle de Rham cohomology of a $D$-dimensional K\"ahler manifold, or more generally of families of algebraic varieties \cite{Griffiths:1968_I,Griffiths:1968_II,Griffiths:1970}. Indeed, in our original discussion of the Weil operator of the torus in \ref{subsec:Weil_SL2} we started from the action of the Hodge star on differential forms. Of course, this requires knowledge of the metric on the torus. For arbitrary K\"ahler manifolds, however, this is in general very difficult. On the other hand, considering instead the action of the Weil operator on the cohomology turns this into a problem of linear algebra and group theory. Indeed, it turns out that one can learn much from studying just the abstract properties of the corresponding Hodge structure, without making any reference to some underlying geometry.

\subsubsection*{Horizontality}

Let us now return to $z$-dependence of the subspaces $H^{p,q}$. To characterize it, one introduces the Hodge filtration $F^p$ as
\begin{equation}
    F^p = H^{D,0}\oplus\cdots\oplus H^{D-p,p}=\bigoplus_{q\geq p} H^{q,D-q}\,.
\end{equation}
This defines a decreasing filtration
\begin{equation}
    F^D\subseteq F^{D-1}\subseteq \cdots \subseteq F^0 = H\,,
\end{equation}
satisfying 
\begin{equation}
H^{p,q}=F^p\cap \overline{F^q}\,,\qquad H = F^p\oplus \overline{F^{D-p+1}}\,. 
\end{equation}
The condition we now impose on how the Hodge structure depends on the parameter $z$ is the direct generalization of \eqref{eq:horizontality_torus}
\begin{equation}
\label{eq:horizontality}
    \frac{\partial}{\partial z}F^p \subseteq F^{p-1}\,,\qquad \frac{\partial}{\partial\bar{ z}}F^p \subseteq F^p\,,
\end{equation}
and is referred to as the horizontality condition. 

\subsubsection*{Asymptotic Behaviour}

Now let us consider the limit $z\rightarrow i\infty$.\footnote{Strictly speaking, one usually starts by considering a variation of Hodge structure that is locally defined over a punctured disk. Then $z$ is introduced as the covering space coordinate, so that the puncture corresponds to the limit $z\rightarrow i\infty$.} According to the nilpotent orbit theorem of W.~Schmid \cite{schmid}, any variation of Hodge structure is \textit{approximately} described by a so-called nilpotent orbit
\begin{equation}
\label{eq:def_nilpotent orbit}
    F^p \approx F^p_{\mathrm{nil}}=e^{z N}F^p_0\,,
\end{equation}
where $N$ is a nilpotent matrix and is referred to as the log-monodromy matrix, since it is associated with a monodromy transformation $z\mapsto z+1$. In general there will be corrections to \eqref{eq:def_nilpotent orbit} which scale as $e^{2\pi i z}$ and are therefore exponentially suppressed in the limit $z\rightarrow i\infty$. The approximation  \eqref{eq:def_nilpotent orbit} is called the nilpotent orbit approximation. Note that in the example of the torus, this relation is exact and there are no exponential corrections. 

The filtration $F^p_0$ is again identified as the limiting filtration. However, as was already apparent in the example of the torus, it will generically not define a Hodge filtration. Instead, it can be shown that, together with the monodromy matrix $N$, it defines a so-called mixed Hodge structure. For the purpose of this introduction, it will not be necessary to delve into the exact details here, but refer the reader to \cite{schmid,CKS,Grimm:2018ohb,Grimm:2018cpv,Grimm:2021ckh} for further details. However, let us comment on its main relevance with regards to the $\mathrm{Sl}(2)$-orbit approximation. 

\subsubsection*{Nilpotent orbit vs. $\mathrm{Sl}(2)$-orbit}

In the example of the torus, we saw that the nilpotent orbit approximation and $\mathrm{Sl}(2)$-orbit approximation coincide, and furthermore describe the full variation of Hodge structure exactly. In general, neither of these statements is true. Indeed, it turns out that for more complicated examples the proposal \eqref{eq:Finfty_torus} does not work. Instead, one first defines a rotated version of $F^p_0$ by
\begin{equation}
\label{eq:Rsplit}
    \tilde{F}^p_0 = e^{\zeta}e^{i\delta} F^p_0\,.
\end{equation}
The operator $\delta$ plays an especially important role, and is referred to as the phase operator. The operators $\delta,\zeta$ can be constructed uniquely from the mixed Hodge structure defined by $N$ and $F^p_0$, see e.g.~\cite{Grimm:2018cpv,Grimm:2021ckh} for further details. Roughly speaking, they are necessary to identify the $\mathfrak{sl}(2)$-triple from which the $\mathrm{Sl}(2)$-orbit will be built. Indeed, given the filtration $\tilde{F}^p_0$ and the monodromy matrix $N$ one can uniquely construct an $\mathfrak{sl}(2)$-triple $\{N^+,N^0,N^-\}$ which acts on the limiting mixed Hodge structure in a particular way. From here the discussion proceeds as for the torus, but using $\tilde{F}^p_0$ instead of $F^p_0$. One defines the boundary Hodge structure by
\begin{equation}
    F^p_\infty = e^{iN}\tilde{F}^p_0\,.
\end{equation}
We stress that it is already a non-trivial statement that $F^p_\infty$ indeed defines a Hodge filtration. In exactly the same manner as we did in the example of the torus, one may associate a boundary charge operator $Q_\infty$ to the boundary Hodge structure $H^{p,q}_\infty$ via the eigenspace decomposition. This in turn defines a boundary Weil operator $C_\infty = (-1)^{Q_\infty}$. Also, using the fact that $p+q=D$, one sees that
\begin{equation}
    C_\infty^2 = (-1)^D\,.
\end{equation}
Indeed, for the torus we have $D=1$ and hence recover the familiar condition $C_\infty^2=-1$.

The $\mathrm{Sl}(2)$-orbit corresponding to $F^p_\infty$ is then similarly defined by
\begin{equation}
\label{eq:def_Sl2_orbit}
    F^p_{\mathrm{Sl}(2)} = e^{xN^-}y^{-\frac{1}{2}N^0} F^p_\infty\,.
\end{equation}
However, there is an important difference with respect to the torus example. There we found that $F^p_{\mathrm{nil}}=F^p_{\mathrm{Sl}(2)}$. In contrast, for more complicated examples this is only true to first order in $y^{-1}$. In this sense, one can think of the $\mathrm{Sl}(2)$-orbit \eqref{eq:def_Sl2_orbit} as the first order approximation in $y^{-1}$ to the full variation of Hodge structure. In general, one instead has
\begin{equation}
\label{eq:Fnil_Finfty}
    F^p_{\mathrm{nil}}=e^{xN^-}\left(1+\frac{g_1}{y}+\frac{g_2}{y^2}+\cdots\right)y^{-\frac{1}{2}N^0}F^p_\infty\,,
\end{equation}
where the operators $g_i$ encode (infinitely many) subleading corrections in powers of $y^{-1}$.
The fact that any variation of Hodge structure is asymptotically described by an $\mathrm{Sl}(2)$-orbit comprises the first part of the famous $\mathrm{Sl}(2)$-orbit theorem \cite{schmid,CKS}. However, perhaps even more strikingly, the second half of the theorem provides an algorithmic procedure to compute the corrections $g_i$ in \eqref{eq:Fnil_Finfty}. The theorem even applies when considering multi-parameter variations of Hodge structure. However, this is extremely non-trivial and goes beyond what is necessary in this work. For a concrete application in the physics literature we refer the reader to \cite{Grimm:2021ikg}, where the full tower of corrections has been explicitly computed and resummed for all one-parameter variations of Hodge structure arising in Calabi--Yau threefolds with a single complex structure modulus.

\subsection{Example: Type $\mathrm{IV}_1$}
\label{subsec:example}

The preceding discussion has been rather abstract, so let us end this section by providing an explicit example of a horizontal $\mathfrak{sl}(2)$-triple in $\mathfrak{sp}(4,\mathbb{R})$ and write down the corresponding solution to the bi-Yang--Baxter model. We also end with some speculative comments regarding the nilpotent orbit approximation discussed in section \ref{subsec:asymp_Hodge_theory}. We refer the reader to \cite{Grimm:2021ikg} for further details on this example.

\subsubsection*{Charge Operator and Boundary Hodge Structure}

The charge operator is given by\footnote{We refer the reader to appendix \ref{app:Rmat} for some details on the algebra $\mathfrak{sp}(4,\mathbb{R})$. }
\begin{equation}
\label{eq:Q_IV1}
   Q_{\infty}=\frac{i}{2}\begin{pmatrix}
   0 & -3 & 0 & 0\\
   1 & 0 & -2 & 0\\
   0 & 2 & 0 & -1\\
   0 & 0 & 3 & 0
   \end{pmatrix}\,.
\end{equation}
For the reader interested in the Hodge-theoretic interpretation, we note that the eigenspace decomposition of this charge operator induces the following weight three Hodge structure
\begin{equation}
    \mathbb{C}^4 = H^{3,0}_\infty\oplus H^{2,1}_\infty\oplus H^{1,2}_\infty\oplus H^{0,3}_\infty\,,
\end{equation}
where
\begin{equation}
    H_\infty^{3,0}=\mathrm{span}_\mathbb{C}\begin{pmatrix}
    1 \\ i \\ -1 \\ -i
    \end{pmatrix}\,,\qquad H_\infty^{2,1}=\mathrm{span}_\mathbb{C}\begin{pmatrix}
    1 \\ \frac{i}{3} \\ \frac{1}{3} \\ i
    \end{pmatrix} \,,
\end{equation}
with $H_\infty^{1,2}$ and $H_\infty^{0,3}$ obtained by complex conjugation. 

\subsubsection*{Horizontal $\mathfrak{sl}(2)$-triple}

The $\mathfrak{sl}(2)$-triple is given by
\begin{equation}
\label{eq:sl2_IV1}
    N^+ = \begin{pmatrix}
    0 & 3 & 0 & 0\\
    0 & 0 & 2 & 0\\
    0 & 0 & 0 & 1\\
    0 & 0 & 0 & 0
    \end{pmatrix}\,,\qquad N^0=\begin{pmatrix}
    3 & 0 & 0 & 0\\
    0 & 1 & 0 & 0\\
    0 & 0 & -1 & 0\\
    0 & 0 & 0 & -3
    \end{pmatrix}\,,\qquad N^-=\begin{pmatrix}
    0 & 0 & 0 & 0\\
    1 & 0 & 0 & 0\\
    0 & 2 & 0 & 0\\
    0 & 0 & 3 & 0
    \end{pmatrix}\,.
\end{equation}
One readily checks that the commutation relations \eqref{eq:Q_commutation} are satisfied. From a Hodge-theoretic perspective, this particular horizontal $\mathfrak{sl}(2)$-triple defines a type $\mathrm{IV}_1$ limiting mixed Hodge structure. Geometrically, it arises for example in the large complex structure limit of a Calabi--Yau threefold with a single complex structure modulus. 

\subsubsection*{Weil Operator}

The Weil operator corresponding to this horizontal $\mathfrak{sl}(2)$-triple via \eqref{eq:def_Weil_general} is explicitly given by
\begin{equation}
\label{eq:Weil_IV1}
    g(x,y) = \frac{1}{y^3}\left(
\begin{array}{cccc}
 -x^3 & 3 x^2 & -3 x & 1 \\
 -x^2 \left(x^2+y^2\right) & 3 x^3+2 x y^2 & -3 x^2-y^2 & x \\
 -x \left(x^2+y^2\right)^2 & \left(x^2+y^2\right) \left(3 x^2+y^2\right) & -3 x^3-2 x y^2 & x^2 \\
 -\left(x^2+y^2\right)^3 & 3 x \left(x^2+y^2\right)^2 & -3 x^2 \left(x^2+y^2\right) & x^3 \\
\end{array}
\right)\,.
\end{equation}
By our general arguments, the operator \eqref{eq:Weil_IV1} together with the $R$-matrix \eqref{eq:def_R_general} provide a solution to the critical $\mathfrak{sp}(4,\mathbb{R})$ bi-Yang--Baxter model, as can be verified by explicit computation. For some further details on how we constructed a full $R$-matrix for $\mathfrak{sp}(4,\mathbb{R})$ we refer the reader to appendix \ref{app:Rmat}, see in particular \eqref{eq:R_IV1}. 

\subsubsection*{Nilpotent Orbit Approximation}

We end this example with an observation regarding the nilpotent orbit approximation discussed in section \ref{subsec:asymp_Hodge_theory}. As a rough summary, it was mentioned that for general variations of Hodge structure, the $\mathrm{Sl}(2)$-orbit approximation \eqref{eq:Weil_IV1} will only provide the first order approximation of the full variation of Hodge structure. To obtain a better approximation one must consider the nilpotent orbit approximation \eqref{eq:Fnil_Finfty} by incorporating an infinite tower of $y^{-1}$ corrections. In \cite{Grimm:2021ikg} this was done explicitly for this particular example. The resulting Weil operator is given by (we have set $x=0$ for simplicity)
\begin{equation*}
\label{eq:Weil_IV1_nilpotent}
    \frac{g(y)}{N(y)} = \begin{pmatrix}
    0 & -9y^2\chi & 0 & 8y^3+\chi\\
    -\frac{3}{2}y(2y^3+\chi)\chi & 0 & -\frac{(2y^3+\chi)(8y^3+\chi)}{2y} & 0\\
    0 & y(8y^6-y^3\chi+2\chi^2) & 0 & -3y^2\chi\\
    -\frac{1}{2}(2y^3+\chi)(8y^6-y^3\chi+2\chi^2) & 0 & -\frac{9}{2}y(2y^3+\chi)\chi & 0
    \end{pmatrix}\,,
\end{equation*}
with 
\begin{equation}
    N(y)=\frac{1}{(4y^3-\chi)(2y^3+\chi)}\,.
\end{equation}
The parameter $\chi$ appearing in $g(y)$ is what drives the corrections. Indeed, in the limit $\chi\rightarrow 0$ one recovers the $\mathrm{Sl}(2)$-orbit result \eqref{eq:Weil_IV1}. From the Hodge-theoretic point of view, it arises precisely in the step \eqref{eq:Rsplit}, where the limiting filtration $F^p_0$ is rotated to $\tilde{F}^p_0$. This step is crucial in order to identify the $\mathfrak{sl}(2)$-triple \eqref{eq:sl2_IV1}. Indeed, the phase operator $\delta$ is given by
\begin{equation}
    \delta = \begin{pmatrix}
    0 & 0 & 0 & 0\\
    0 & 0 & 0 & 0\\
    0 & 0 & 0 & 0\\
    \chi & 0& 0 & 0
    \end{pmatrix}\,.
\end{equation}
In the geometric setting, when this boundary Hodge structure arises in the large complex structure regime of a Calabi--Yau threefold $Y_3$, the parameter $\chi$ is proportional to the Euler characteristic of $Y_3$. 

The curious reader may wonder whether also the expression \eqref{eq:Weil_IV1_nilpotent} solves the bi-Yang--Baxter model, since it at least does so in the $\chi\rightarrow 0$ limit. If one naively takes the same $R$ matrix \eqref{eq:R_IV1} as was used for the $\mathrm{Sl}(2)$-orbit approximation, one will find that it does not provide a solution. However, it is an interesting possibility that an appropriate dependence of $R$ on the parameter $\chi$ alleviates this issue, thus promoting the full nilpotent orbit approximation to a solution of the associated bi-Yang--Baxter model. If this is indeed the case, this would imply a remarkable connection between the $R$ matrix and the phase operator $\delta$. 

At present, there is no concrete evidence that this will indeed be the case. There are, however, two indications. The first is that sometimes the nilpotent orbit and $\mathrm{Sl}(2)$-orbit are related in a rather simple manner. Indeed, it was found in \cite{Grimm:2021ikg} for the type $\mathrm{I}_1$ and $\mathrm{II}_0$ boundaries, that after appropriately resumming the corrections in the nilpotent orbit approximation, the two are related by a simple coordinate shift $y\mapsto y+y_0$. Of course, such a shift will not spoil the solution. The second indication is that in an earlier work \cite{Grimm:2021idu} it was shown that in fact the full Weil operator (hence also the nilpotent orbit approximation) solves the equations of motion of the $\lambda$-deformed $G/G$ model. Therefore, there is already an established relationship between objects appearing in Hodge theory and deformations of integrable non-linear $\sigma$-models. It appears plausible, then, that this relation runs deeper and also applies to the bi-Yang--Baxter model beyond just the $\mathrm{Sl}(2)$-orbit approximation. This is, however, still rather speculative and we hope to return to this question in future work. 

\section{Conclusions}

In this work, we have presented a new class of solutions to the critical bi-Yang--Baxter model. Although this model is integrable, writing down explicit solutions to its equations of motion is still a non-trivial task. It was shown in \cite{Schepers:2020ehn} that for the simplest case of the $\mathrm{SU}(2)$ bi-Yang--Baxter model it is possible to obtain a subset of special solutions by appropriately deforming the so-called uniton solutions of the principal chiral model found in the seminal work \cite{Uhlenbeck:1989}. However, this procedure relies on manipulations that explicitly involve the target space coordinates and is therefore not feasible to extend to higher-dimensional groups. Instead, we have introduced a class of solutions that does not require a coordinate-dependent formulation, but which is instead described purely in terms of group-theoretic objects. As a result, our solutions are very general and can be written down for groups of arbitrary dimension.  

More specifically, the solutions we have presented correspond to $\mathrm{Sl}(2)$-orbits and were originally motivated by considering the action of the Hodge star operator on the middle de Rham cohomology of a two-torus. Indeed, on the one hand, we have argued that the latter explicitly solves the critical $\mathrm{SL}(2,\mathbb{R})$ bi-Yang--Baxter model. In fact, it was shown that it can be identified with a special complex uniton solution of the $\mathrm{SU}(2)$ bi-Yang--Baxter model at the critical point $\zeta=\eta$, where the two deformation parameters are equal and the symmetry of the model enhances. On the other hand, the Hodge star operator, viewed as an operator on the middle cohomology, is an example of an important Hodge-theoretic object called the Weil operator. In the case of the two-torus, the Weil operator is exactly described by an $\mathrm{Sl}(2)$-orbit. Crucially, the concept of an $\mathrm{Sl}(2)$-orbit is much more general and can be written down for higher-dimensional groups. Practically, the relevant information required to construct such orbits is fully encoded in terms of a horizontal $\mathfrak{sl}(2)$-triple inside the algebra of the group under consideration. The main result of this work is now the following: using the properties of the horizontal $\mathfrak{sl}(2)$-triple, we (1) identify a subset of $R$-matrices compatible with the triple, i.e.~satisfying \eqref{eq:def_R_general}, and we (2) show that the $\mathrm{Sl}(2)$-orbit \eqref{eq:def_Weil_general} corresponding to the triple solves the equations of motion of the critical bi-Yang--Baxter model defined by this subset of $R$-matrices. The solutions thus constructed have finite action and satisfy $C^2=(-1)^D$ and therefore provide a generalization of the complex uniton solutions of the $\mathrm{SU}(2)$-model. Importantly, since horizontal $\mathfrak{sl}(2)$-triples have been classified in the mathematics literature \cite{Robles:2015,Kerr2017} the corresponding generalized unitons and $R$-matrices are also classified. 

A first question, which is of immediate interest, is whether our proposed solutions can be generalized further. Here the underlying Hodge theory provides a very natural candidate for such a generalization. Indeed, as was explained in section \ref{subsec:asymp_Hodge_theory}, one can think of the $\mathrm{Sl}(2)$-orbit as the first-order approximation to an arbitrary variation of Hodge structure, which is valid near the boundary of the parameter space over which the variation is defined. A natural question, therefore, is whether possible corrections to the $\mathrm{Sl}(2)$-orbit would spoil the solution to the bi-Yang--Baxter model. This is a very non-trivial question, as the computation of such corrections is already rather involved and constitutes the second part of the celebrated $\mathrm{Sl}(2)$-orbit theorem \cite{schmid,CKS}. The resulting orbit, including all these corrections, is referred to as the nilpotent orbit. Despite the complexity of the algorithm, in examples the corrections can be explicitly computed, see e.g.~\cite{Grimm:2021ikg}, and hence the fate of our proposed solutions can be tested. We expect that the $R$-matrix will play an especially important role and that, in order for the corrections to not spoil the solution, it might need to be constructed using the additional information associated to a  
nilpotent orbit. We hope to return to this question in future work. 

Another question that arises is how much the concepts in Hodge theory underlie the study of solutions to integrable non-linear $\sigma$-models in general. Indeed, it is now clear that the Weil operator plays a special role in this regard, as it can be used to construct solutions to the critical bi-Yang--Baxter model as shown in this work, and also for the $\lambda$-deformed $G/G$ model as shown in our previous work \cite{Grimm:2021idu}. Furthermore, via Poisson--Lie T-duality one expects that the solutions can be mapped to the dual models. It would be interesting to investigate how far our proposed solutions extend across the duality web of integrable field theories. Of perhaps even greater importance is to establish the exact relevance of integrability in this regard. We expect that there is a special type of integrability that makes models amenable to a treatment with Hodge-theoretic techniques. 

The connection of Hodge theory to integrable models also gives another way 
of highlighting how tame geometry enters physics. It has recently been suggested that tameness is a general property of physical theories, such as effective theories compatible with quantum gravity \cite{Grimm:2021vpn} as well as conformal field theories and certain classes of quantum field theories \cite{Douglas:2022ynw,DouglasPartII}.
Tame geometry is currently 
a very active field of mathematics, which is, in part, due to some striking developments that reveal the underlying tame structures in Hodge theory, see \cite{Bakker:2017,Bakker:2020,Bakker:2021uqw}. Indeed, it has been shown that the period map and the Weil operator of a general variation of Hodge structure are tame. Applied to the bi-Yang--Baxter or $\lambda$-deformed $G/G$ model, it implies that some of its solutions are tame functions. This requires 
a careful specification of the global properties of the solutions. Furthermore,   
we expect that the integrable models that we considered here can also be studied further at the quantum level, see e.g.~\cite{Bazhanov:2018xzh,Schepers:2020ehn}. They can thus serve as a testing ground for the general tameness conjectures put forward in~\cite{DouglasPartII} that claim the tameness of the correlation functions varying with the parameters of the theory and over space-time. 

As a last point, we would like to ponder the possibility of coupling the bi-Yang--Baxter model to gravity, e.g.~along the lines of JT-gravity \cite{Teitelboim:1983ux,Jackiw:1984je}. Indeed, one of the original motivations of our work was to establish a proper $\sigma$-model formulation of Hodge theoretic objects such as the period map and Weil operator, as was suggested in \cite{Grimm:2020cda}, see also \cite{Cecotti:2020uek}. However, an object that has remained rather mysterious in this regard is the so-called Weil--Petersson metric. While this metric arises very naturally in the context of string compactifications, its precise significance from a Hodge-theoretic point of view is not much explored. It was suggested in \cite{Grimm:2020cda} that the Weil--Petersson metric emerges as the classical solution to a non-linear $\sigma$-model coupled to gravity. It would be interesting to see if this can be achieved using the bi-Yang--Baxter model. This would also open the possibility to develop a holographic description of the two-dimensional gravity theory as envisioned in \cite{Grimm:2020cda}. 

\subsubsection*{Acknowledgements}

We would like to thank Falk Hassler, Damian van de Heisteeg, Dirk Schuricht, Daniel Thompson and Mick van Vliet for useful discussions and comments. This research is supported, in part, by the Dutch Research Council (NWO) via a
Start-Up grant and a Vici grant.

\pagebreak
\appendix 

\section{Overview of Formulae}
\label{app:computations}

In this section we have collected some formulae that are used in the computations of section \ref{subsec:sl2_orbits}. 

\subsubsection*{Action of $\mathrm{Ad}_h$ and $\mathrm{Ad}_{h^{-1}}$}
In the $\mathrm{Sl}(2)$-orbit approximation, the period map is given by
\begin{equation}
    h = e^{xN^-}y^{-\frac{1}{2}N^0}\,.
\end{equation}
Using the commutation relations \eqref{eq:sl2_algebra}, it follows by direct computation that
\begin{align}
\label{eq:Adh_N+}
    \mathrm{Ad}_h N^+ &= \frac{1}{y}N^+-\frac{x}{y}N^0-\frac{x^2}{y}N^-\,,\\
\label{eq:Adh_N0}
    \mathrm{Ad}_h N^0 &=N^0+2xN^-\,,\\
\label{eq:Adh_N-}
    \mathrm{Ad}_h N^- &= yN^-\,.
\end{align}
In a similar fashion, one finds
\begin{align}
\label{eq:Adhinv_N+}
    \mathrm{Ad}_{h^{-1}}N^+ &= yN^++xN^0-\frac{x^2}{y}N^-\,,\\
\label{eq:Adhinv_N0}
    \mathrm{Ad}_{h^{-1}}N^0 &= N^0-\frac{2x}{y}N^-\,,\\
\label{eq:Adhinv_N-}
    \mathrm{Ad}_{h^{-1}}N^- &= \frac{1}{y}N^-\,.
\end{align}

\subsubsection*{Action of $(-1)^{\mathrm{ad}\,Q_{\infty}}$}
It will be convenient to denote
\begin{equation}
    \mathcal{O}^\dagger = -(-1)^{\mathrm{ad}\,Q_{\infty}}\mathcal{O}\,.
\end{equation}
Then using the commutation relations \eqref{eq:sl2_algebra} and \eqref{eq:Q_commutation}, one finds
\begin{align}
\label{eq:dagger_N+}
    \left(N^+\right)^\dagger &= N^-\,,\\
\label{eq:dagger_N0}
    \left(N^0\right)^\dagger &= N^0\,,\\
\label{eq:dagger_N-}
    \left(N^-\right)^\dagger &= N^+\,.
\end{align}

\subsubsection*{Action of $\mathrm{Ad}_g$}
We recall that 
\begin{equation}
    g = h(-1)^{Q_{\infty}}h^{-1}\,,
\end{equation}
so that
\begin{equation}
    \mathrm{Ad}_g = -\mathrm{Ad}_h\circ \dagger\circ\mathrm{Ad}_{h^{-1}}\,.
\end{equation}
In order to compute the action of $\mathrm{Ad}_g$ on the $\mathfrak{sl}(2)$-triple, we can use our earlier results \eqref{eq:Adh_N+}-\eqref{eq:Adhinv_N-} as well as \eqref{eq:dagger_N+}-\eqref{eq:dagger_N-}.  Then one finds
\begin{align}
\label{eq:Adg_N+}
    \mathrm{Ad}_g N^+ &= \frac{1}{y^2}\left[x^2 N^+-x(x^2+y^2)N^0-(x^2+y^2)^2N^- \right]\,,\\
\label{eq:Adg_N0}
    \mathrm{Ad}_g N^0 &= \frac{2}{y^2}\left[x N^+-\left(x^2+\frac{y^2}{2}\right)N^0-x(x^2+y^2)N^- \right]\,,\\
\label{eq:Adg_N-}
    \mathrm{Ad}_g N^-&=-\frac{1}{y^2}\left[N^+-x N^0-x^2 N^- \right]\,.
\end{align}
Note that since $g^2=\pm 1$, we also have $\mathrm{Ad}_g=\mathrm{Ad}_{g^{-1}}$.

\section{$R$-matrix for type $\mathrm{IV}_1$}
\label{app:Rmat}

In this section we explicitly write down the $R$-matrix used in section \ref{subsec:example}, where the type $\mathrm{IV}_1$ Weil operator is discussed. Recall that $\mathfrak{sp}(4,\mathbb{R})$ consists of real $4\times 4$ matrices $X$ satisfying
\begin{equation}
    X^T\cdot S+S\cdot X=0\,,
\end{equation}
where $S$ is a non-singular skew-symmetric matrix and $T$ denotes the transpose. In this particular example we have made the following (non-standard) choice
\begin{equation}
    S = \begin{pmatrix}
    0 & 0 & 0 & -1\\
    0 & 0 & 3 & 0\\
    0 & -3 & 0 & 0\\
    1 & 0 & 0 & 0
    \end{pmatrix}\,.
\end{equation}
A basis of $\mathfrak{sp}(4,\mathbb{R})$ is given by
\begin{equation}
    T_k = \frac{1}{k!}\left(\mathrm{ad}\,N^+\right)^k \left(N^-\right)^3\,,\qquad k=0,\ldots, 6\,,
\end{equation}
together with
\begin{equation}
    T_7 = N^+\,,\qquad T_8 = N^0\,,\qquad T_{9}=N^-\,,
\end{equation}
with $N^+,N^0,N^-$ given in \eqref{eq:sl2_IV1}. Note that this particular embedding of $\mathfrak{sl}(2,\mathbb{R})$ into $\mathfrak{sp}(4,\mathbb{R})$ corresponds to the $\mathbf{10}=\mathbf{3}\oplus\mathbf{7}$ representation. 

With the charge operator $Q_\infty$ in \eqref{eq:Q_IV1} at hand, it is straightforward to define an $R$-matrix that furthermore satisfies the condition \eqref{eq:def_R_general}. Indeed, one simply demands that $R$ commutes with $Q_\infty$ and $L_0$ and acts as $\pm c$ on the eigenvectors of $\mathrm{ad}\,Q_\infty$ with positive/negative eigenvalues, respectively. In other words, we put
\begin{equation}
    R\,Q_\infty=0\,,\qquad R\,L_0 = 0\,,\qquad R\,\mathcal{O}_q = c\, \mathrm{sign}(q)\mathcal{O}_q\,,\qquad [Q_\infty, \mathcal{O}_q] = q\,\mathcal{O}_q\,.
\end{equation}
For this particular example, these equations have a unique solution if one furthermore demands that $R$ is anti-symmetric, as is required in the bi-Yang--Baxter model. Of course, the resulting solution is essentially the Drinfel'd--Jimbo solution.\footnote{In principle, there is another Cartan generator besides $Q_\infty$, but one can check that taking this into account does not change the solution in this case.} In the basis of $\{T_k\}$ it reads
\begin{equation}
\label{eq:R_IV1}
    R=ic \left(
\begin{array}{cccccccccc}
 0 & \frac{5}{16} & 0 & \frac{1}{16} & 0 & \frac{1}{16} & 0 & 0 & 0 & 0 \\
 -\frac{15}{8} & 0 & \frac{3}{8} & 0 & \frac{1}{8} & 0 & \frac{3}{8} & 0 & 0 & 0 \\
 0 & -\frac{15}{16} & 0 & \frac{9}{16} & 0 & \frac{5}{16} & 0 & 0 & 0 & 0 \\
 -\frac{5}{4} & 0 & -\frac{3}{4} & 0 & \frac{3}{4} & 0 & \frac{5}{4} & 0 & 0 & 0 \\
 0 & -\frac{5}{16} & 0 & -\frac{9}{16} & 0 & \frac{15}{16} & 0 & 0 & 0 & 0 \\
 -\frac{3}{8} & 0 & -\frac{1}{8} & 0 & -\frac{3}{8} & 0 & \frac{15}{8} & 0 & 0 & 0 \\
 0 & -\frac{1}{16} & 0 & -\frac{1}{16} & 0 & -\frac{5}{16} & 0 & 0 & 0 & 0 \\
 0 & 0 & 0 & 0 & 0 & 0 & 0 & 0 & \frac{1}{2} & 0 \\
 0 & 0 & 0 & 0 & 0 & 0 & 0 & -1 & 0 & -1 \\
 0 & 0 & 0 & 0 & 0 & 0 & 0 & 0 & \frac{1}{2} & 0 \\
\end{array}
\right)\,.
\end{equation}
One may verify that \eqref{eq:R_IV1} satisfies the modified classical Yang--Baxter equation \eqref{eq:YBE}. Furthermore, one sees that when $c=i$ the $R$-matrix is indeed a real endomorphism of $\mathfrak{sp}(4,\mathbb{R})$. By examining the $3\times 3$ block on the bottom right-hand side, one verifies that it acts on the real $\mathfrak{sl}(2)$-triple as required by \eqref{eq:def_R_general}. In contrast, it acts on the remaining $7\times 7$ block in a more complicated fashion.  

\bibliographystyle{JHEP}
\bibliography{references}

\end{document}